\documentclass[11 pt]{article}
\usepackage{amssymb,latexsym,amsmath,amsfonts}
\usepackage{subfigure,color}
\usepackage{graphicx, epsf}
\usepackage[active]{srcltx}
\usepackage[active]{srcltx}
\numberwithin{equation}{section}

\usepackage{amssymb,latexsym,amsmath,amsfonts}
\usepackage{graphicx, epsf, subfigure}
\usepackage{epsfig}

\def\ep{\varepsilon}

\def\kcb{\bar{k}_c}

\def\bfrho{\mbox{\boldmath$\rho$}}
\def\bfpsi{\mbox{\boldmath$\psi$}}

\begin{document}

\title{Turing pattern formation in the Brusselator system with nonlinear diffusion}

\author{G. Gambino\footnote{Department of Mathematics, University of Palermo, Italy, gaetana@math.unipa.it}$\;$
M.C. Lombardo\footnote{Department of Mathematics, University of Palermo, Italy, lombardo@math.unipa.it}$\;$
M. Sammartino\footnote{Department of Mathematics, University of Palermo, Italy, marco@math.unipa.it}$\;$
V. Sciacca\footnote{Department of Mathematics, University of Palermo, Italy, sciacca@math.unipa.it}}

\maketitle

\begin{abstract}
In this work we investigate the effect of density dependent nonlinear diffusion on pattern
formation in the Brusselator system.
Through linear stability analysis of the basic solution we determine the Turing and the oscillatory
instability boundaries.
A comparison with the classical linear diffusion shows how nonlinear diffusion favors the
occurrence of Turing pattern formation.
We study the process of pattern formation both in 1D and 2D spatial domains.
Through a  weakly nonlinear multiple scales analysis we derive the equations
for the amplitude of the stationary patterns. The analysis of the amplitude equations shows
the occurrence of a number of different
phenomena, including stable supercritical and subcritical Turing patterns with multiple branches
of stable solutions leading to hysteresis.
Moreover we consider traveling patterning waves: when the domain size is
large, the pattern forms sequentially and traveling wavefronts are the precursors to patterning.
We derive the Ginzburg-Landau equation and describe the traveling front enveloping a pattern
which invades the domain.
We show the emergence of radially symmetric target patterns, and through a matching procedure we construct
the  outer amplitude equation and the inner core solution.\end{abstract}

%\begin{keyword}
%Generalized Camassa-Holm Equations, Traveling waves, Homoclinic and Heteroclinic Orbits
%\end{keyword}

%\end{frontmatter}

\section{Introduction}\label{Sec1}

The aim of this work is to describe the Turing
pattern formation for the following reaction-diffusion system, introduced in \cite{KH10}:
\begin{equation}\label{or_syst}
\begin{split}
\displaystyle\frac{\partial U}{\partial \tau}&=D_u \displaystyle\frac{\partial}{\partial \zeta} \left(\left(\frac{U}{u_0}\right)^m\frac{\partial U}{\partial \zeta}\right)+\Gamma\left(a-(b+1)U+U^2V\right),\\
\displaystyle\frac{\partial V}{\partial \tau}&=D_v \displaystyle\frac{\partial}{\partial \zeta} \left(\left(\frac{V}{v_0}\right)^n\frac{\partial V}{\partial \zeta}\right)+\Gamma\left(bU-U^2V\right).
\end{split}
\end{equation}
Here  $U(\zeta,\tau)$ and $V(\zeta,\tau)$, with $\zeta\in [0,l]$,
represent the concentrations of two chemical species,
the activator and the inhibitor respectively;  $a$ and $b$ are positive constants;
the constant $\Gamma >0$ represents the strength of the reaction terms or, alternatively,  modulates the
size of the domain.

Equations \eqref{or_syst} belong to the class of reaction-diffusion systems with nonlinear diffusion.
The nonlinear density-dependent diffusion present in  \eqref{or_syst}
is such  that, when $m,n>0$,
the species tend to diffuse faster (when $U>u_0$ and $V>v_0$) or slower (when $U<u_0$ and $V<v_0$)
than predicted by the linear classical diffusion.
The coefficients $D_u,D_v>0$ are the classical diffusion coefficients and
the nonnegative $u_0$ and $v_0$ are threshold concentrations which measure the strength of the interactions
between the individuals of the same species.
At  microscopic level this kind of diffusion term can be interpreted as the result of the interaction between
random walkers representing the individuals of the system; whereas classical diffusion corresponds
to the case of independent random walks.
In particular, the dynamics is sub-diffusive as the mean square displacement of the particles $\left<(\Delta \zeta)^2\right> \sim \tau^\alpha$, with $\alpha=1/(2+m)<1$ (for the $V$ species $\alpha=1/(2+n)$), see the discussion in \cite{Kumar09}.

The reaction mechanism is chosen as in the Brusselator autocatalytic system.
This system is a model used to capture the qualitative behavior of  cross activator-inhibitor chemical reactions:
to this class belong some autocatalytic reactions such as  ferrocyanide-iodate-sulphite reaction, chlorite-iodide-malonic acid (CIMA) reaction, arsenite-iodate reaction, and  many enzyme catalytic reactions.
For a review on the rich spatial and temporal dynamics shown by cubic-autocatalytic reaction-diffusion
systems, see e.g. \cite{GS,EP}.

%Usually pattern forming chemical systems have been investigated considering constant diffusion
%coefficients; this is a simplifying assumption, not suitable for many
%reaction-diffusion systems, especially in the realm of biological or
%geological systems with heterogeneity.
%Even the first experimental demonstration  of the formation of Turing structures \cite{LE91}, obtained for
%the CIMA reaction with hydrogel,  was subsequently revealed  to be strongly affected by the binding of
%the activator (iodide) with a slowly diffusing substrate (starch);
%this phenomenon, effectively reducing the diffusion coefficient of the activator,
%was in fact the ultimate cause of the Turing instability, see \cite{HS88}.
%It was later shown \cite{RR04} that the whole system can be appropriately modeled introducing
%concentration dependent diffusion coefficients.
\vskip 5cm

In general one can say that models like \eqref{or_syst} are believed to be relevant for
autocatalytic chemical reactions occurring: a) on a binding hydrogel substrate \cite{LGV96,NN-UM93};
b) on surfaces \cite{IE95,RW01,RR04}, like cellular membranes, or in phenomena of industrial interest
like surface electrodeposition \cite{BLMS12} or metal catalysis \cite{Hil02,LGWWSW03};
c) on porous media\cite{ASAST13,ZHMO01,ZHMOL02}.

Recently reaction-diffusion models with concentration-dependent diffusion coefficients have
attracted considerable attention in many different fields \cite{GLS09,GLS12,MRW11,TLP10,Jun10,Gal12,R-BT12,Sherratt2000,TV07,DCL02,MPT05,CJ07};
however the effect of the anomalous diffusivity on Turing pattern of chemical and biochemical
systems is not yet fully  investigated, due to the presence of complicated reaction terms and
related difficulties even in the linear stability analysis.
As exceptions here we mention \cite{Mal88}, where it was shown that the introduction of
concentration dependence of the diffusion coefficient, due to the ionic character of the reactants,
sharpens the features of the pattern, resulting in an increase of the
chemical gradients of the chemicals; or the paper \cite{RW01}, where pattern formation in the Gray-Scott model of
excitable media with diffusion coefficient linearly dependent on the concentration was studied;
the mechanism responsible for the formation of the pattern was however different from Turing bifurcation;
and the recent \cite{LHPLZS12} where the authors consider and explore, numerically,
the Lengyel-Epstein model with local concentration-dependent diffusivity.
More specifically for  the Brusselator system,  recently, have appeared several papers considering
pattern formation in presence of cross-diffusion \cite{KH11,ZKHH13,ZVE11}
and superdiffusion due to the fractional Laplacian \cite{GMV08,TMV09,TBMV11}.

Regarding the system \eqref{or_syst}, in \cite{KH10} the authors derived the conditions for Turing instability
and showed that, differently from the standard linear diffusion case, the destabilization of the
constant steady state, occurs even if the diffusion constant $D_v$ of the inhibitor is smaller or
equal to the diffusion constant $D_u$ of the activator.
In this paper we reconsider the linear stability analysis of the system \eqref{or_syst} taking into account
the fact that the Brusselator kinetics also supports Hopf bifurcation: even though the steady state is Turing unstable,
whether Turing patterns form depends on the mutual location of the Hopf and the Turing instability boundaries.
This will be analyzed in Section \ref{Sec2}, where  the Turing and Hopf stability boundaries
will be obtained in terms of three key system parameters.
This will clarify the role of nonlinear diffusion in the formation of the pattern.
In Section \ref{Sec3} we shall perform the weakly nonlinear analysis near the onset of the Turing instability.
The amplitude equation will be derived both for stationary pattern (Stuart-Landau amplitude equation) and
spatially modulated pattern (real Ginzburg-Landau amplitude equation).
Moreover, we shall derive the quintic Stuart-Landau equation which describes the phenomenon of hysteresis
occurring in the case when the bifurcation is subcritical.
Numerical simulations are performed to corroborate the predictions coming from
the weakly nonlinear analysis.
In Section \ref{Sec4} we shall focus on pattern formation in a 2D domain.
Rolls and squares, which arise when the homogeneous steady state bifurcates at a simple eigenvalue,
and mixed-mode patterns, which emerge when the eigenvalue is double and different modes interact,
will be shown.
Particular mixed-mode patterns are the hexagonal patterns, which appear when a resonance condition holds.
The evolution system for the amplitudes of the patterns in each case will be given and discussed.
The emergence of axisymmetric target patterns will then be shown and an asymptotic matching procedure will be employed to derive the appropriate amplitude equation.

We finally believe that, in the context of the previously mentioned  phenomena of surface chemical reactions
and chemical instabilities in porous media, the  model \eqref{or_syst} and the mathematical analysis
presented in this paper, could be of support for the quantitative prediction of the observed dynamics and
for the design of more focused experiments.
For example, as discussed in \cite{KH10}, the analysis presented here (and in \cite{KH10}) should motivate
and solicit the realization of experiments devoted to the investigation of dissipative structures in
open chemical reactors where porous media are employed.

\section{Linear stability analysis\label{Sec2}}
\setcounter{figure}{0}
\setcounter{equation}{0}
In analogy with \cite{GMV08}, we rescale \eqref{or_syst}, using $U=u^*u,\ V=v^*v,\ \tau=t,\ \zeta=x^*x\ $,
 where:
\begin{equation}\label{nodim}
\begin{split}
u^*=&\,\left(\frac{(m+1)D_vu_o^m}{(n+1)D_uv_0^n}\right)^{\textstyle\frac{1}{m+n+2}},\quad v^*=\frac{1}{u^*},\\
x^*=&\,\sqrt{\frac{D_v}{(n+1)v_0^nu^{*(n+2)}}},
\end{split}
\end{equation}
to obtain:
\begin{equation}\label{syst}
\begin{split}
\frac{\partial u}{\partial t}=&\,\frac{\partial^2}{\partial x^2} u^{m+1}+\Gamma(Q-(b+1)u+u^2v),\\
\frac{\partial v}{\partial t}=&\,\frac{1}{\eta^2}\frac{\partial^2}{\partial x^2}v^{n+1}+\frac{\Gamma}{\eta^2}(bu-u^2v),
\end{split}
\end{equation}
having defined:
\begin{equation}\label{Qeta}
Q=a\eta\qquad{\rm and}\qquad \eta=1/u^*.
\end{equation}
The system \eqref{syst} is supplemented with initial data and Neumann boundary conditions.
%(we are interested in self-organizing patterns and the chosen BCs impose the weakest
%constraint on pattern formation).
The only nontrivial homogeneous stationary solution admitted by the system \eqref{syst}
is $(\bar{u},\bar{v})\equiv (Q, b/Q)$.
Through linear stability analysis one gets the following dispersion relation which gives the growth
rate $\sigma$ as a function of the wavenumber $k$:
\begin{equation}\label{2.4}
\sigma^2+g(k^2)\sigma+h(k^2)=0
\; ,
\end{equation}
where:
\[\begin{split}
g(k^2)=&\; k^2 \,{\rm tr}(D)-\Gamma \,{\rm tr}(K) ,\\
h(k^2)=&\;{\rm det}(D)k^4+\Gamma q k^2+\Gamma^2 {\rm det}(K),\end{split}\]
with:
\begin{equation}\label{2.2}
\begin{split}
K=&\,\left(\begin{array}{cc}
b-1  & Q^2\\
-\displaystyle\frac{b}{\eta^2} & -\displaystyle\left(\frac{Q}{\eta}\right)^2
\end{array}\right),\\
D=&\,\left(\begin{array}{cc}
(m+1)Q^m & 0\\
0 & \displaystyle\frac{n+1}{\eta^2}\left(\frac{b}{Q}\right)^n
\end{array}\right)\!,
\end{split}
\end{equation}
and $q=-K_{11}D_{22}-K_{22}D_{11}$.
The steady state $(\bar{u},\bar{v})$ can lose its stability both via Hopf and Turing bifurcation.
Oscillatory instability occurs when $g(k^2)=0$ and $h(k^2)>0$.
The minimum values of $b$ and $k$ for which $g(k^2)=0$ are:
\begin{equation}\label{hopf}
b_c^H=1+\frac{Q^2}{\eta^2}\, \qquad  k=0,
\end{equation}
and for $b>b_c^H$ a spatially homogeneous oscillatory mode emerges.
The neutral stability Turing boundary corresponds to $h(k^2)=0$,
which has a single minimum $(k_c^2, b^c)$ attained when:
\begin{equation}\label{kc2}
k_c^2=-\Gamma\frac{(1-b)(n+1)\displaystyle\left(\frac{b}{Q}\right)^n+(m+1)Q^{m+2}}{2\, (m+1)(n+1)Q^{m-n}b^n},
\end{equation}
which requires $q<0$, therefore $b>1$ is a necessary condition for Turing instability. Then, the Turing bifurcation value $b=b^c$ is obtained by imposing $q^2 -4{\rm det}(D){\rm det}(K)=0$ (under the condition $q<0$), which leads to solve:
\begin{equation}\label{turing}
\left\{\begin{split}
&\,\left((n+1)\left(\frac{b}{Q}\right)^n(1-b)+(m+1)Q^{m+2}\right)^2\\&\hskip1.8cm-4(m+1)(n+1)Q^{m-n+2}b^n=0, \\
&\,Q^{m+n+2}<\frac{(n+1)}{(m+1)}b^n(b-1).
\end{split}\right.
\end{equation}
%
%For $b>b_c$ a finite band of unstable wavenumbers for which $Re(\sigma)>0$ stays in between the roots of
%$h(k^2)$ (which are proportional to $\Gamma$) and Turing instability occurs.
From \eqref{kc2} and \eqref{turing} one can easily see that the
first mode to lose stability, that we shall denote with $k_c$,  and the bifurcation value $b^c$
do not depend explicitly on $\eta$.
However being impossible to give an explicit expression, we have evaluated $k_c$ and $b^c$ numerically.

\begin{figure}[h]
\begin{center}
{\epsfxsize=2.3 in \epsfysize=2.1 in\epsfbox{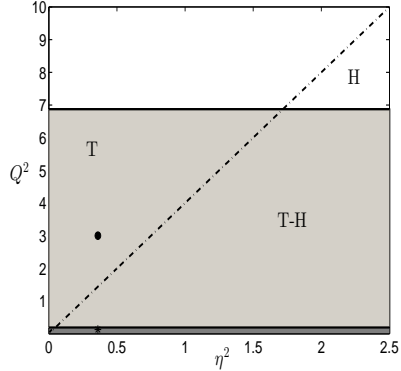}}
\end{center}
\caption{\label{fig2_1} Hopf (dash-dotted line) and Turing (solid line) stability boundaries in the two
dimensional slice $\{b=\mbox{const}\}$, with $b>b_c$.
The instability Turing region is shadowed in grey: the supercritical region in light grey,
the subcritical region in dark grey (see Sections \ref{Subsec3.1} and \ref{Subsec3.2} below).
The pure Hopf region, the pure Turing region and the region in which both Turing and Hopf instability
occur are labeled with H, T and T-H respectively.
The parameters are chosen as $m=n=1$ and $b=11>b_c=10.74$.}
\end{figure}

\begin{figure}[h]
\begin{center}
\subfigure[] {\epsfxsize=2.0 in \epsfbox{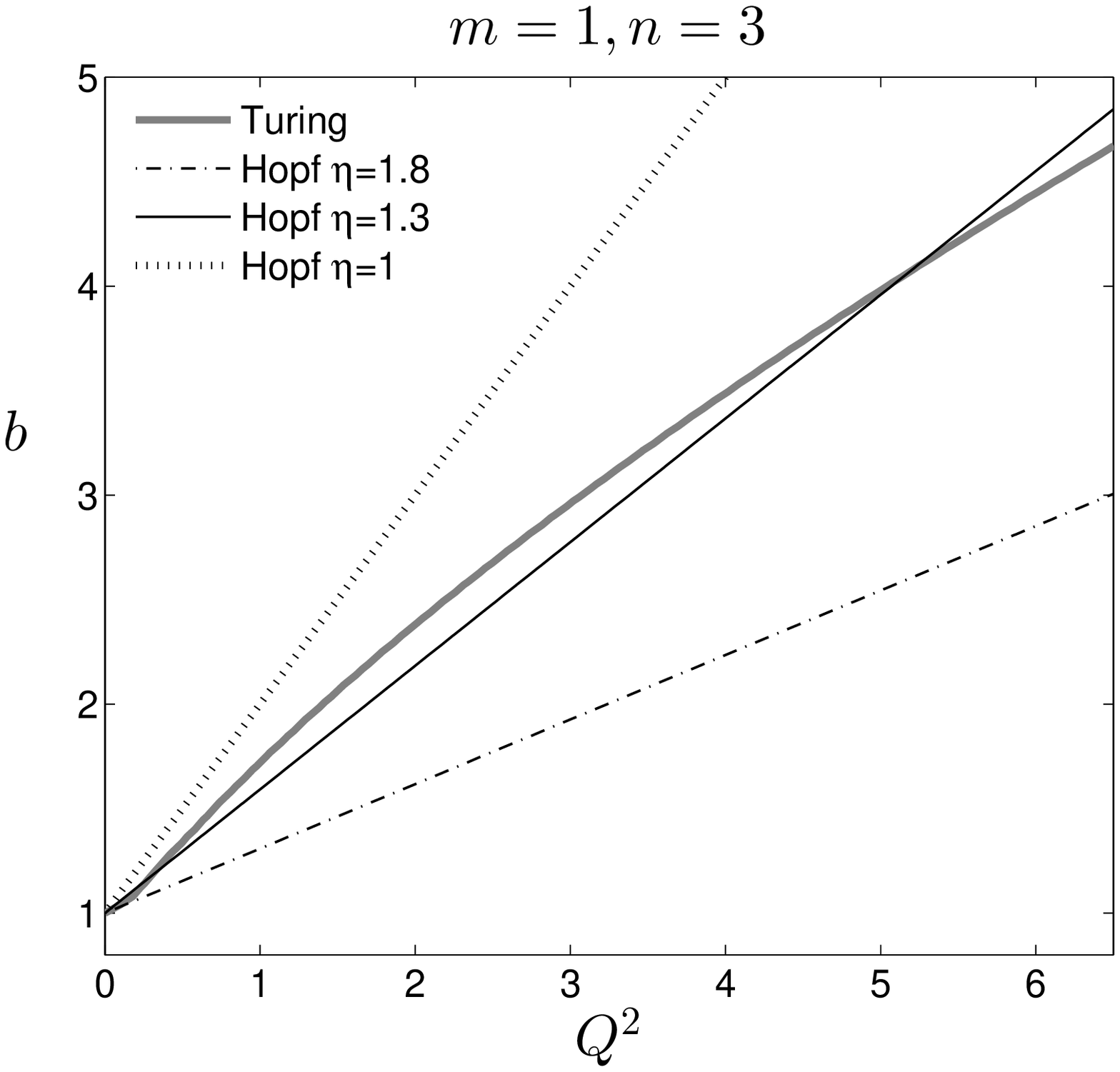}}
\subfigure[] {\epsfxsize=2.0 in \epsfbox{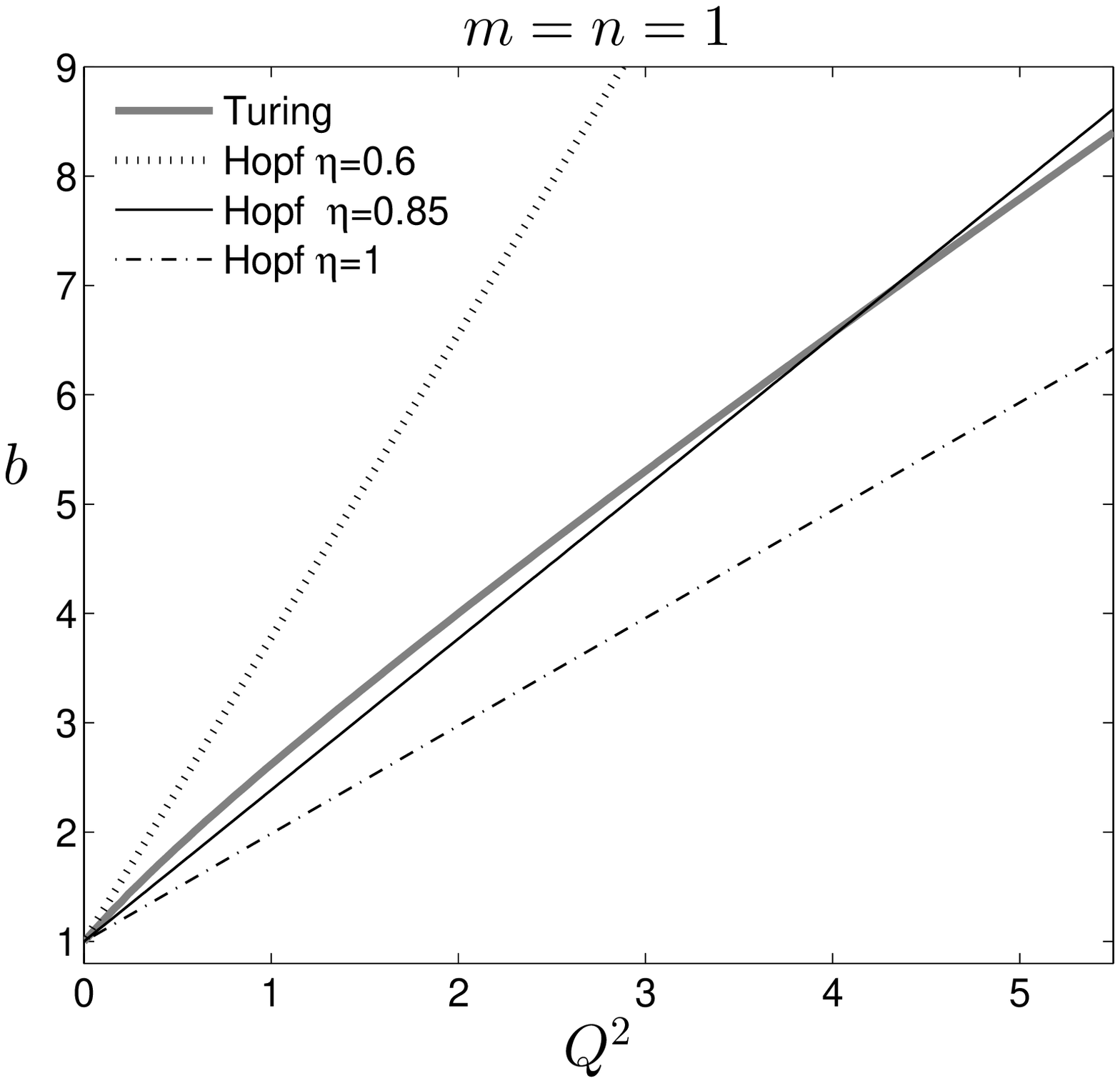}}
\subfigure[] {\epsfxsize=2.0 in \epsfbox{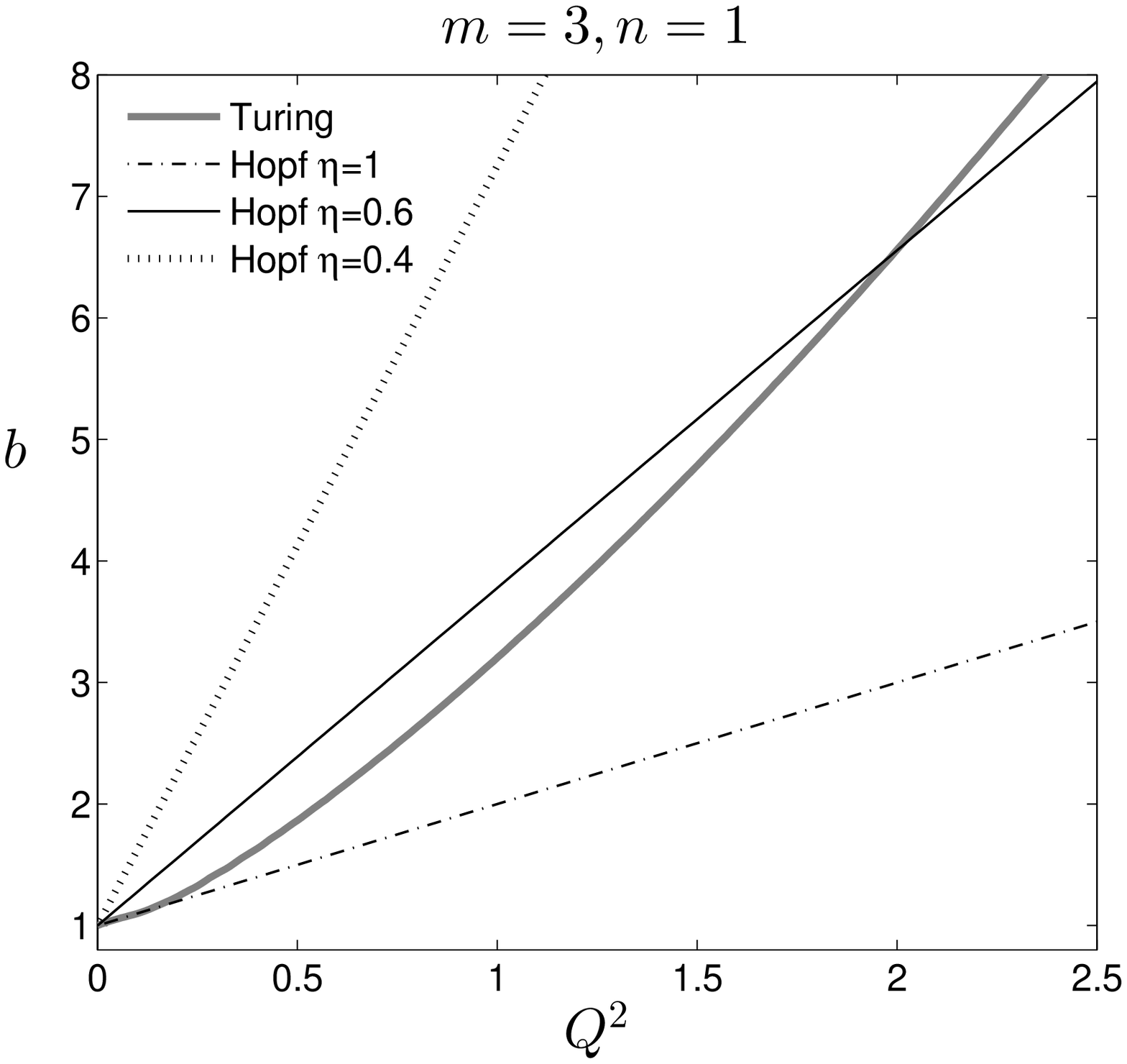}}
\end{center}
\caption{\label{fig2_2} In Figs (a), (b), (c) we represent the Hopf and Turing stability boundaries for different values of $\eta$. The instability regions stay above the lines.}
\end{figure}

In Fig.\ref{fig2_1} we show the Turing and Hopf instability regions in the parameter space $(\eta^2, Q^2)$
for a fixed value $b>b^c$.
In the region marked with TH there is a competition between the two instabilities; which one, between the Turing
and the Hopf instability develops, depends on the locations of the respective instability boundaries:
as $b$ increases, if $b_c^H>b^c$, Turing instability occurs prior to the oscillatory instability.

In Fig.\ref{fig2_2}(a), (b) and (c) we report the Hopf and Turing instability boundaries for different values
of $\eta$ and different values of the coefficients $m$ and $n$ expressing the nonlinearity of the diffusion.
We observe that, while the Turing instability region does not depend on $\eta$,
the Turing pattern region decreases by increasing $\eta$ due to the fact that
$b_c^H$ decreases with $\eta$.
Comparing the three Figs \ref{fig2_2}(a)-\ref{fig2_2}(c) it is interesting to notice that, taking larger values
of $m$ with respect to $n$, the shape of the Turing boundaries changes concavity; which leads to
the fact that Turing patterns develop for large enough values of $Q$ when $m$ is smaller than $n$
(see Fig.\ref{fig2_2}(a)),
and for small enough values of $Q$ when $m$ is larger than $n$ (see Fig.\ref{fig2_2}(c)) .

\begin{figure}[h]
\begin{center}
%\subfigure[] {\epsfxsize=2.3 in \epsfbox{mnUguali_1.eps}}
%\subfigure[] {\epsfxsize=2.3 in \epsfbox{mnUguali_2.eps}}
\subfigure[] {\epsfxsize=2.5 in \epsfbox{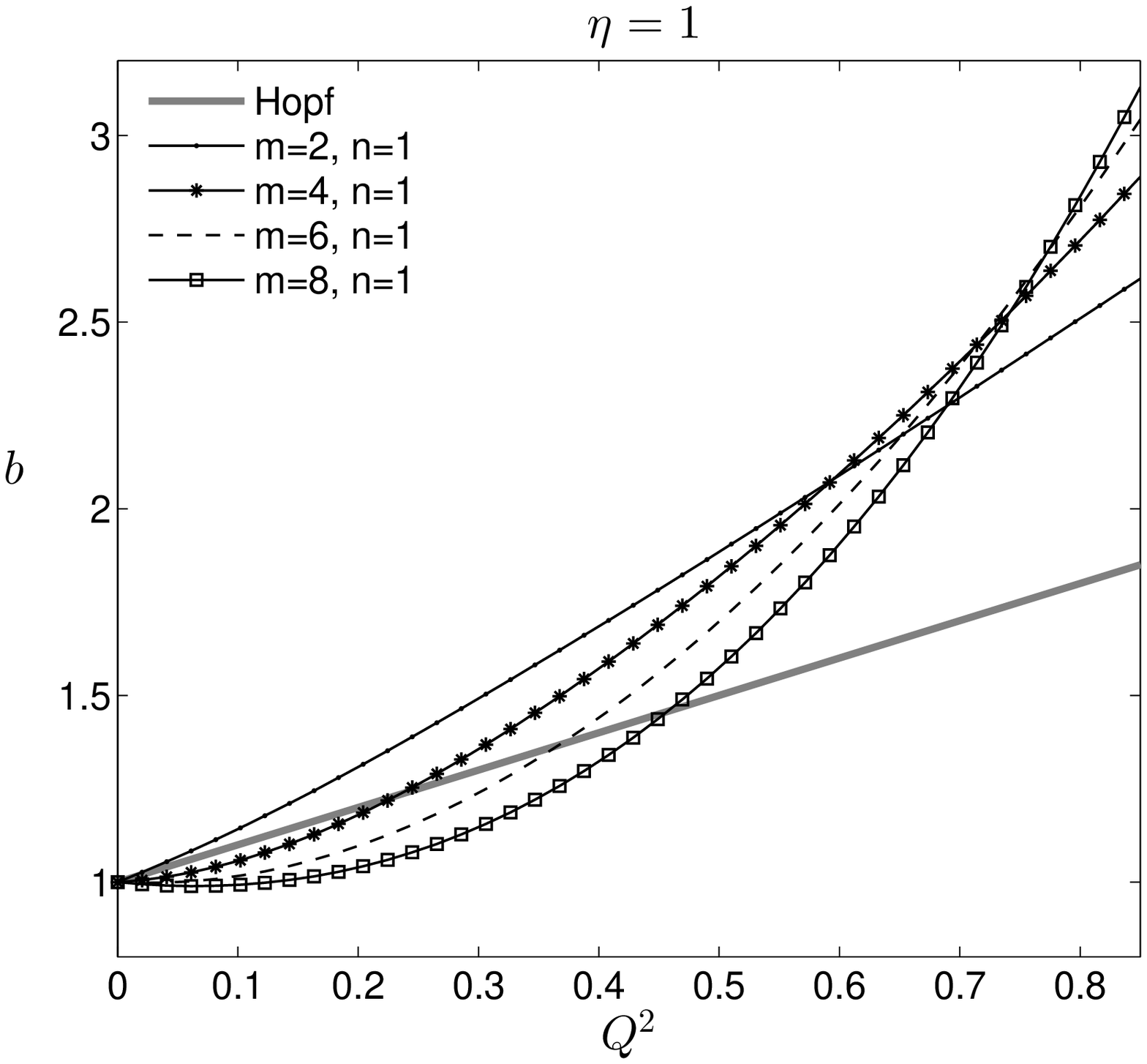}}
\subfigure[] {\epsfxsize=2.5 in \epsfbox{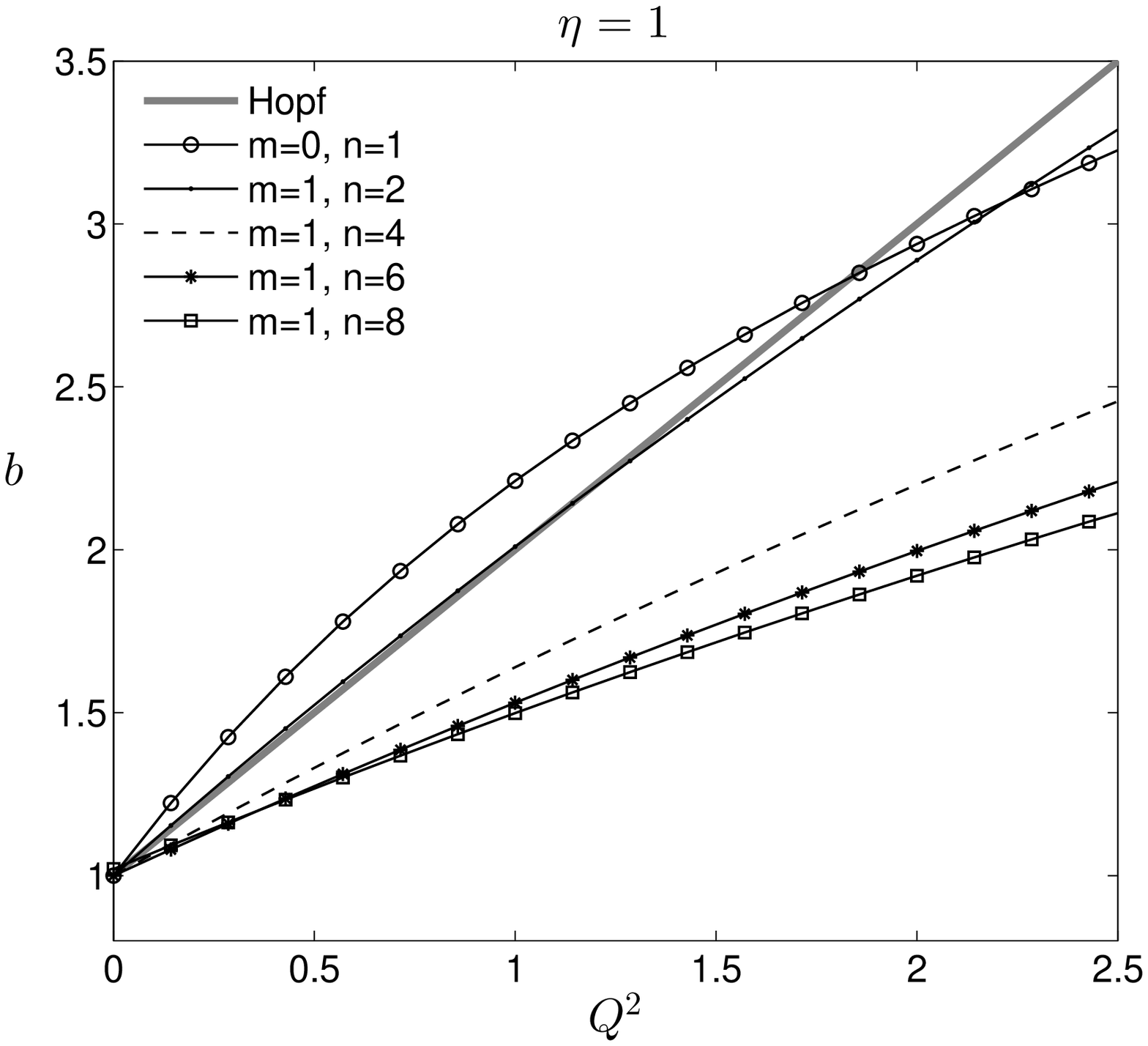}}
\end{center}
\caption{\label{fig2_3} Hopf and Turing stability boundaries varying $m$ and $n$. The instability regions stay above the lines.}
\end{figure}

The effect of  nonlinear diffusion is to make easier the formation of Turing pattern with respect to
the case of classical diffusion (i.e. when $m=n=0$).
This is apparent from Fig.\ref{fig2_3} where we have reported the stability boundaries for
different values of $m$ and $n$ and keeping $\eta=1$.

%%%when  (classical diffusion) the pattern forms
%%%only when the diffusion coefficient of the inhibitor $D_v$ is sufficiently greater than that of the activator
%%%$D_u$, which means $\eta<<1$;
%%%we recover the same result only when $m\geq n$ and $n\leq 1$, see Fig.\ref{stab_bound2}-(a);
%%%in all the other cases when $m,n\neq0$ Turing pattern arises even for $\eta\geq1$.
%%%In particular, when $m=n>1$, the Turing instability boundary is a convex increasing function and intersects
%%%the Hopf boundary, in the plane $(Q^2, b)$, at some $Q^2=Q^2_c$ for which the thresholds $b_c$ and $b_c^H$
%%%coincide even for $\eta\geq1$ (and correspondingly $D_v<D_u$). This means that, when $b>b_c$, Turing pattern
%%%forms $\forall Q^2\in (0,Q^2_c)$  and the Turing region increases with $m$ and $n$, see Fig.\ref{stab_bound2}-(b).
%%%When $m>n$, the Turing instability boundary is still a convex increasing function, which never intersects
%%%the Hopf line till $n<1$, therefore no Turing pattern forms. Once $n\geq1$ an intersection point arises and
%%%the pattern forming region increases with $m$, see Fig.\ref{stab_bound2}-(c).
%%%Finally, when $m<n$ the behaviour is opposite: the Turing instability curve is a concave increasing function,
%%%which intersects the Hopf line at some $Q^2=Q^2_c$ and the Turing pattern forms $\forall Q^2>Q^2_c$.
%%%This intersection disappears by increasing $n$ and the Turing boundary is all below the Hopf line, therefore
%%%the pattern region increases with $n$, see Fig.\ref{stab_bound2}-(d).

\section{Weakly nonlinear analysis\label{Sec3}}
\setcounter{equation}{0}
%
%We use the method of multiple scales to determine the amplitude equation of the pattern close to the instability onset.
We introduce the control parameter $\varepsilon$, representing the dimensionless distance from the
threshold and defined as $\varepsilon^2=(b-b^c)/b^c$;
then we recast the original system \eqref{or_syst} in the following form:
\begin{equation}\label{3.1}
\partial_t \textbf{w}= \mathcal{L}^{b} \textbf{w}+
\mathcal{NL}^{b} \textbf{w},\qquad
\textbf{w}\equiv\left(\begin{array}{c}{u-\bar{u}}\\
{v-\bar{v}}\end{array}\right) \; ,
\end{equation}
where the linear operator $\mathcal{L}^{b}=\Gamma \, K^b + D^{b} \nabla^2$ results from
the linearization of the kinetics and of the diffusion term around the steady state $(\bar{u},\bar{v})$,
the matrix $K^b$ and $D^{b}$ being  given in \eqref{2.2};
here the dependence on the bifurcation parameter ${b}$ is made explicit for notational convenience.
The nonlinear operator $\mathcal{NL}^{b}$ represents the nonlinear remaining terms.
%%
%\begin{equation}\nonumber\label{opNL}
%\mathcal{NL}(\textbf{w})=\Gamma \left(
%\begin{array}{c}
%2\bar{u}uv+\bar{v}u^2+u^2v \\
%-\displaystyle\frac{1}{\eta^2}(2\bar{u}uv+\bar{v}u^2+u^2v)
%\end{array}\right)+\nabla^2\left(
%\begin{array}{c}
%\sum_{k=2}^{m+1}\left(\begin{array}{c}m+1\\k\end{array}\right)\bar{u}^{m+1-k}u^k\\
%\sum_{k=2}^{n+1}\left(\begin{array}{c}n+1\\k\end{array}\right)\bar{v}^{n+1-k}v^k
%\end{array}\right).
%\end{equation}
%%
Moreover we expand $\textbf{w}$, the time $t$ and the bifurcation parameter $b$ as:
\begin{equation}\label{3.3}
\begin{split}
\textbf{w}=&\,\varepsilon \, \textbf{w}_1 +\varepsilon^2\,
 \textbf{w}_2+\dots\quad
 t=t+\varepsilon\, T_1+ \varepsilon^2\, T_2+\dots\\
b=&\,b^c+\varepsilon b^{(1)}+\varepsilon^2 b^{(2)}+\dots
 \end{split}
\end{equation}
Substituting the above expansions into (\ref{3.1})
and collecting the terms at each order in $\varepsilon$, we obtain
the following sequence of equations for the $\textbf{w}_i$:

$O(\varepsilon):$
\begin{equation}\label{sequence_1}
\mathcal{L}^{b^c} {\bf w}_1=\mathbf{0},
\end{equation}
$\ \,O(\varepsilon^2):$
\begin{equation}\label{sequence_2}
\mathcal{L}^{b^c} {\bf w}_2=\mathbf{F},
\end{equation}
$\ \,O(\varepsilon^3):$
\begin{equation}\label{sequence_3}
\mathcal{L}^{b^c} {\bf w}_3=\mathbf{G},
\end{equation}
where:
\[
\begin{split}
\mathbf{F}=&\,\frac{\partial {\bf
w}_1}{\partial T_1}-D^{(1)}\nabla^2\left(\begin{array}{c} u_1^2\\\left(v_1+\displaystyle\frac{b^{(1)}}{Q}\right)^2\end{array}\right)
+
\left(\begin{array}{cc}
-b^{(1)} & 0\\
b^{(1)} & 0
\end{array}
\right){\bf w}_1\\
+&\,\left(2Qu_1v_1+\frac{b^c}{Q}u_1^2\right)\underline{\mathfrak{1}},
\end{split}
\]
\[
\begin{split}
\mathbf{G}=&\,\frac{\partial {\bf
w}_1}{\partial T_2}+\frac{\partial {\bf
w}_2}{\partial T_1}-D^{(2)}\nabla^2
\left(\begin{array}{c}
u_1^3\\
\left(v_1+\displaystyle\frac{b^{(1)}}{Q}\right)^3
\end{array}
\right)
\\
-&\,
2D^{(1)}\nabla^2
\left(\begin{array}{c}
u_1u_2\\
\left(v_1+\displaystyle\frac{b^{(1)}}{Q}\right)\left(v_2+\displaystyle\frac{b^{(2)}}{Q}\right)
\end{array}
\right)
\\
+&\,2\left(Q(u_1v_2+u_2v_1)+\frac{b^c u_1u_2}{Q}+\frac{u_1^2}{2}\left(v_1+\displaystyle\frac{b^{(1)}}{Q}\right)\right)\underline{\mathfrak{1}}\\
+&\,
\left(\begin{array}{cc}
-b^{(1)} & 0\\
b^{(1)} & 0
\end{array}
\right){\bf w}_2+
\left(\begin{array}{cc}
-b^{(2)} & 0\\
b^{(2)} & 0
\end{array}
\right){\bf w}_1,
\end{split}
\]
and:
\begin{equation}\nonumber
\underline{\mathfrak{1}}=\left(\begin{array}{c} -1\\1\end{array}\right),
\end{equation}
\begin{equation}\nonumber
D^{(1)}=\left(
\begin{array}{cc}
\displaystyle\frac{m(m+1)}{2}
Q^{m-1} & 0\\
0 & \displaystyle\frac{n(n+1)}{2\eta^2}\left(\displaystyle\frac{b^c}{Q}\right)^{n-1}
\end{array}
\right),
\end{equation}
\begin{equation}\nonumber
D^{(2)}=\left(
\begin{array}{cc}
\displaystyle\frac{m(m^2-1)}{6}
Q^{m-2} & 0\\
0 & \displaystyle\frac{n(n^2-1)}{6\eta^2}\left(\frac{b^c}{Q}\right)^{n-2}
\end{array}
\right).
\end{equation}

The solution to the linear problem
\eqref{sequence_1}, satisfying the Neumann boundary conditions, is given by:
\begin{equation} \label{3.5}
{\bf w}_1=A(T_1, T_2) \bfrho \, \cos(\bar{k}_c x) \;,
\end{equation}
with $\bfrho \in {\rm Ker}(\Gamma K^{b^c}-\bar{k}_c^2D^{b^c})$ and $\bar{k}_c$ is the first
admissible unstable mode.
%%%%%%%%%%%%%%%%%%%%%%%%%%%%%%%%%%%Inserimento%%%%%%%%%%%%%%%%%%%%%%%%%%%%%%%%%%%%%%%%%%%%
\begin{figure*}[!]
\begin{center}
\subfigure[] {\epsfxsize=2.0 in \epsfysize=2 in \epsfbox{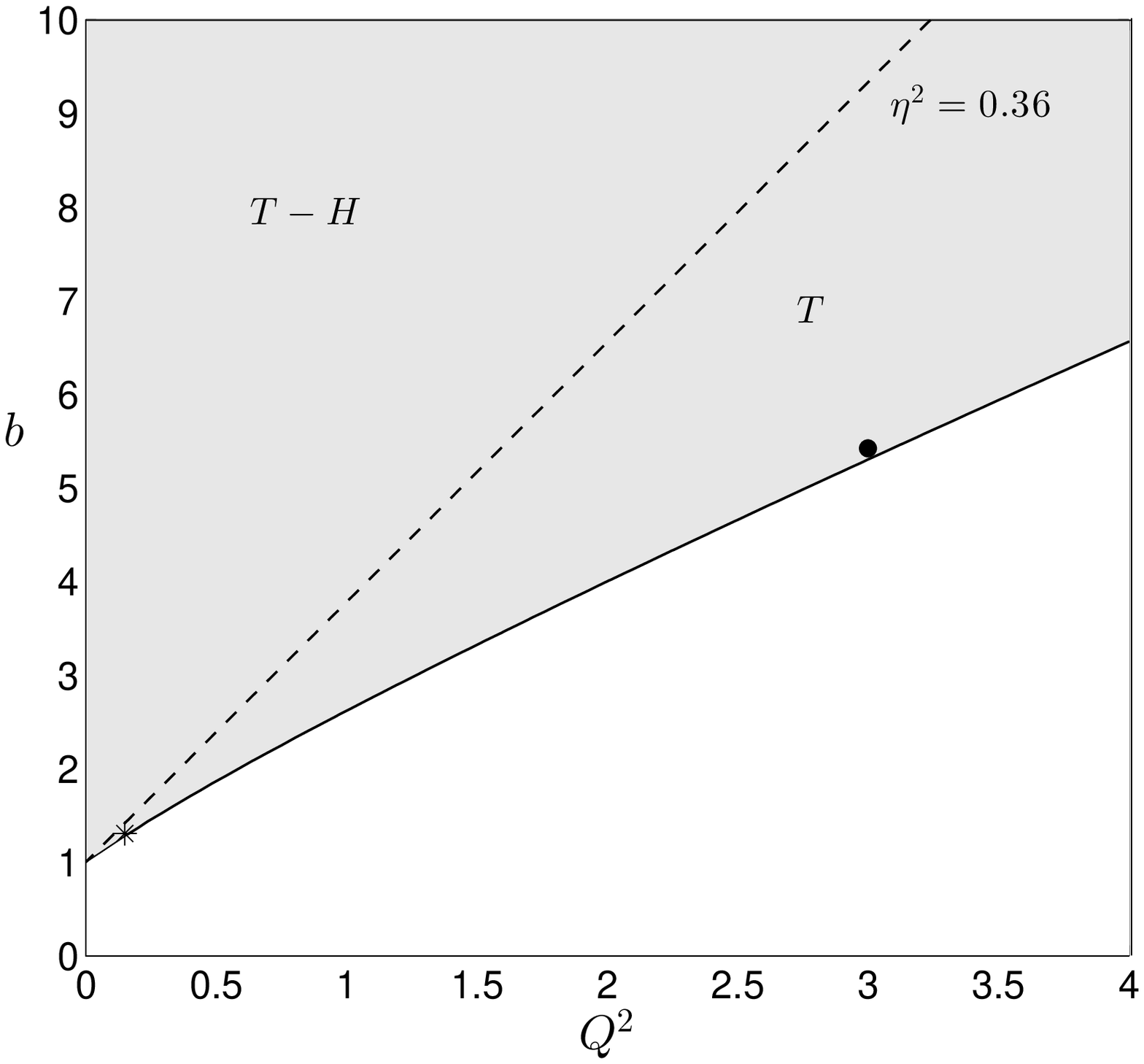}}
\subfigure[] {\epsfxsize=2.0 in \epsfysize=2 in \epsfbox{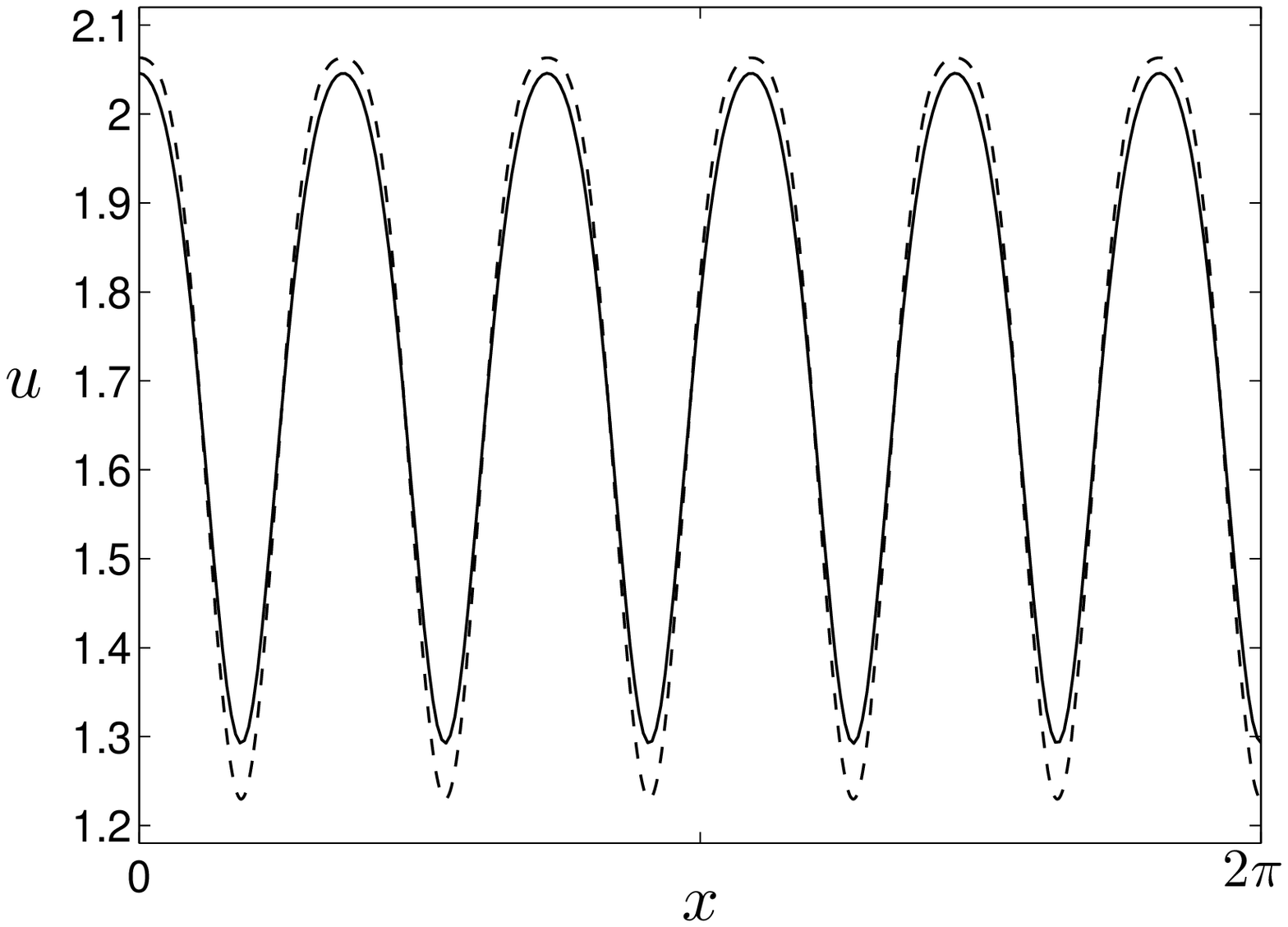}}
\subfigure[] {\epsfxsize=2.0 in \epsfysize=2 in\epsfbox{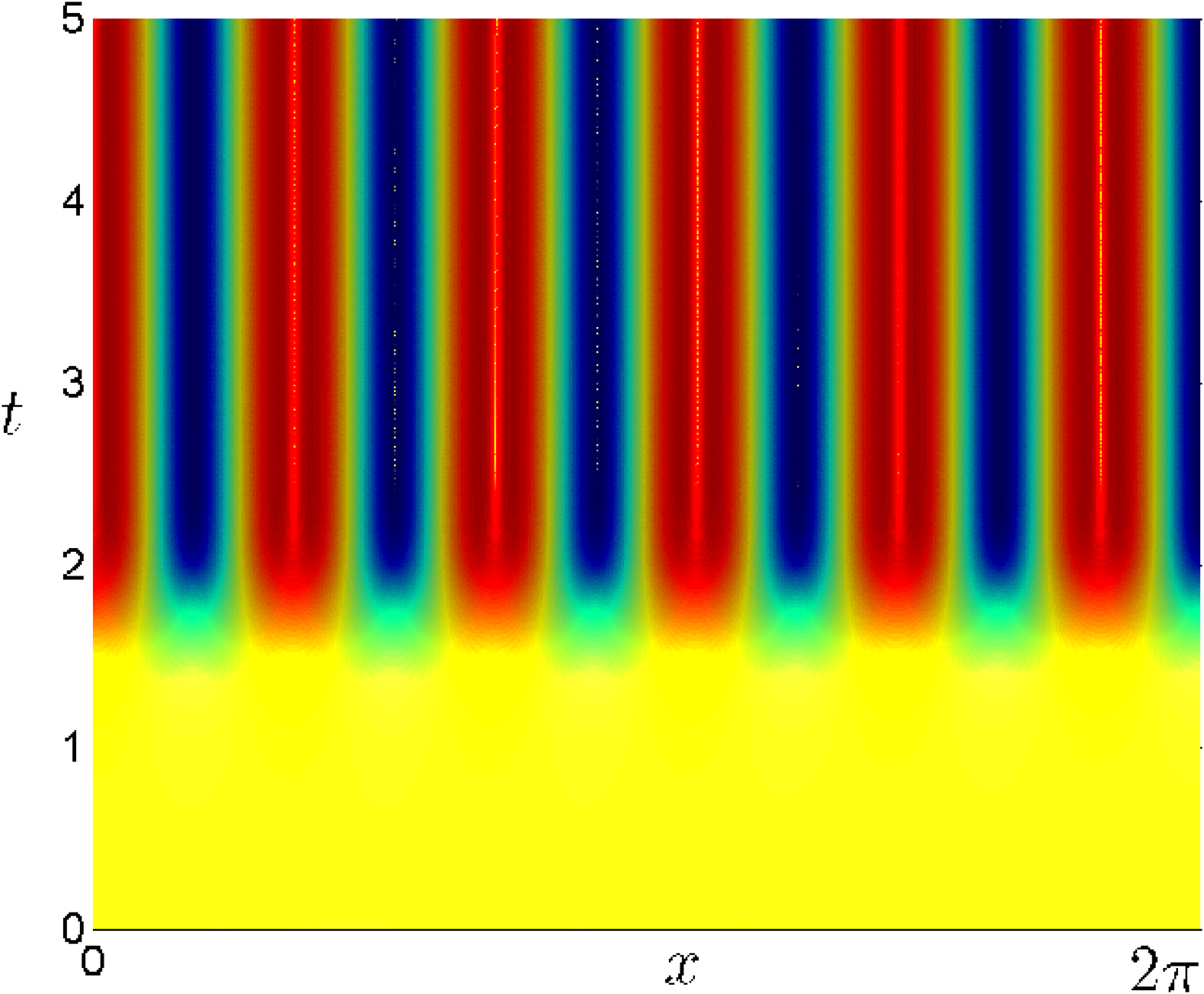}}
\end{center}
\caption{\label{fig3_1} (a) The Turing instability (solid line) occurs prior to oscillatory instability (dashed line). (b) Comparison between the weakly nonlinear solution (dotted line) and the numerical solution of \eqref{or_syst} (solid line). (c) Pattern evolution in the space-time plane. The
parameters are chosen in the supercritical region $\Gamma= 80$, $m=n=1$, $Q^2=3$, $\eta^2=0.36$, $\varepsilon= 0.1$, corresponding to the point marked with the dark circle both in Fig.\ref{fig2_1} and in Fig.\ref{fig3_1}(a). With
this choice of the parameters one has $b^c\approx 5.3028$, while $\bar{k}_c= 5.5$.
}
\end{figure*}
Once substituted the solution \eqref{3.5} in \eqref{sequence_2}, the vector ${\bf F}$ is made orthogonal
to the kernel of the adjoint of $\mathcal{L}^{b^c}$ simply
by imposing $T_1=0$ and $b^{(1)}=0$. The solution of \eqref{sequence_2} can therefore be obtained right away and substituted into the
linear problem \eqref{sequence_3} at order $\varepsilon^3$.
The vector $\mathbf{G}$ has the following expression:
\begin{equation}\label{G}
{\bf G}=\left(\displaystyle\frac{d A}{d
T}\bfrho+A {\bf G}_1^{(1)}+A^3 {\bf G}_1^{(3)}
\right)\cos(\bar{k}_c x)+{\bf G}^* ,
\end{equation}
where $T=T_2$, ${\bf G}_1^j, j=1,3$ and ${\bf G}^*$ (which contains automatically orthogonal terms) depend on the parameters
of the original system \eqref{or_syst} and their explicit expression is here omitted as it is too
cumbersome.
The elimination of secular terms in the equation \eqref{sequence_3} results in the following Stuart-Landau equation
for the amplitude $A(T)$:
\begin{equation}\label{3.11}
\frac{d A}{d T}= \sigma A -L A^3,
\end{equation}
where $\sigma$ and $L$ are explicitly computed in terms of the
system parameters:
\begin{equation}\label{sigmaL}
\sigma=-\frac{<{\bf G}_1^{(1)}, \bfpsi>}{<\bfrho,
\bfpsi>},\qquad  L=\frac{<{\bf G}_1^{(3)}, \bfpsi>}{<\bfrho,
\bfpsi>}
\end{equation}
and ${\bfpsi} \in
Ker\left\{\left(\Gamma K^{b^c} -\bar{k}_c^2D^{b^c}
\right)^\dag\right\}$.
Experimental evidence shows that, when the domain size is large, the pattern sequentially forms and
travelling wavefronts are the precursors to patterning.
Therefore the amplitude of the pattern is also modulated in space.
The slow spatial scale $X=\varepsilon x$ can be easily obtained from the
linear analysis. At $O(\varepsilon)$ we still recover the linear problem $\mathcal{L}^{b^c}
{\bf w}_1=\mathbf{0}$ and the solution is as in \eqref{3.1}, but $A\equiv A(X,T)$ also depends on the slow spatial scale $X$.
Dealing with the equations at $O(\varepsilon^2)$ and $O(\varepsilon^3)$ as before, we obtain
the following Ginzburg-Landau equation for the amplitude $A(X,T)$:
\begin{equation}\label{3.29}
\frac{\partial A}{\partial T}=\nu \frac{\partial^2 A}{\partial
X^2}+ \sigma A -L A^3\,,
\end{equation}
where:
\begin{equation}
\nu=-\frac{<2\bar{k}_c D^{b^c}\mathbf{w}_{21}+D^{b^c}\bfrho.
\bfpsi>}{<\bfrho, \bfpsi>}.
\end{equation}
Here ${\bf w}_{21}$ is the solution of the following linear system:
\begin{equation}\label{lin_sisGL}
(\Gamma K^{b^c}-\bar{k}_c^2 D^{b^c}) {\bf w}_{21}=-2 \bar{k}_c
D^{b^c}\bfrho
\end{equation}
and $\sigma$ and $L$ are
the same as in formulas \eqref{sigmaL}.
\subsection{The supercritical case\label{Subsec3.1}}

In the pattern-forming region, the growth rate coefficient $\sigma$ is always positive.
Therefore two different qualitative dynamics of the Stuart-Landau equation \eqref{3.11} can be identified based on the sign of the coefficient $L$: $L>0$ corresponds to the supercritical case and $L<0$ to the subcritical case (see Fig.\ref{fig2_1}).
In the supercritical case the Stuart-Landau equation admits the stable equilibrium
solution $A_\infty=\sqrt{{\sigma}/{L}}$, which corresponds to the
asymptotic value of the amplitude $A$ of the pattern.
In Fig.\ref{fig3_1}(b), we show the comparison between the stationary solution predicted by
the weakly nonlinear analysis up to $O(\varepsilon^2)$ and the pattern solution computed solving numerically the system \eqref{or_syst} starting from a random periodic perturbation
of the constant state.
In all the tests we have performed we have verified that the distance,
evaluated in the $L^1$ norm, between the weakly nonlinear approximation
and the numerical solution of the system is $O(\varepsilon^3)$.
%
%%%%%%%%%%%%%%%%%%Inserimento%%%%%%%%%%%%%%%%%%%%%%%%%%%%%%%%%%
\begin{figure*}[!]
\begin{center}
{\epsfxsize=13cm\epsfysize=5cm\epsfbox{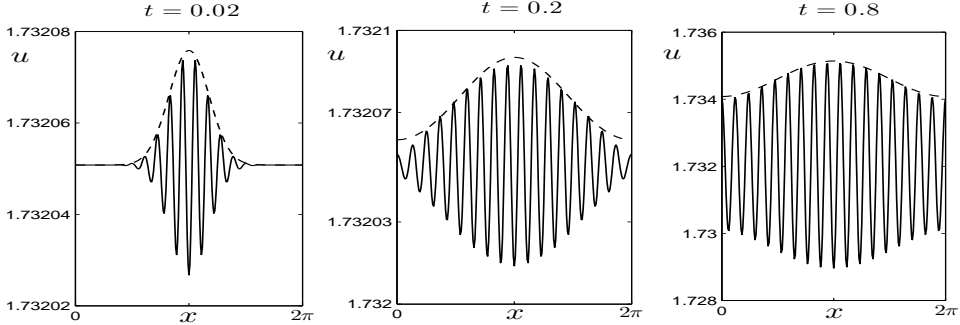}}
\end{center}
\caption{\label{fig3_2}The equilibrium solution is perturbed at the center of the spatial interval.
The pattern forms as a modulated wave (solid line). The dashed line is a numerical
solution of the Ginzburg-Landau equation \eqref{3.29}. The parameters are chosen as in Fig.\ref{fig3_1}, with $\Gamma=800$. Here $\varepsilon=10^{-3}$. }
\end{figure*}
%%%%%%%%%%%%%%%%%%%%%%%%%%%%%%%%%%%%%%%%%%%%%%%%%%%%%%%%%%%%%%%%%%%%%%%%%
Let us consider the same parameter set as in Fig.\ref{fig3_1}, except $\Gamma=800$ larger by a factor $10$, which is equivalent to have a spatial domain larger by a factor $\sqrt{10}$. Once one perturbs the equilibrium solution at the center
of the spatial domain, the pattern propagating as a
wave is observed. In Fig.\ref{fig3_2} it is shown how the Ginzburg-Landau equation \eqref{3.29} is able
to capture the envelope evolution and the progressing of the
pattern as a wave.

\subsection{The subcritical case\label{Subsec3.2}}
%
%%%%%%%%%%%%%%%%%%%%%%%%%%%%%%%%%%%Inserimento%%%%%%%%%%%%%%%%%%%%%%%%%%%%%%%%%%%%%%%%%%%%
\begin{figure*}[!]
\begin{center}
\subfigure[]{\epsfxsize=6.5cm\epsfysize=5cm\epsfbox{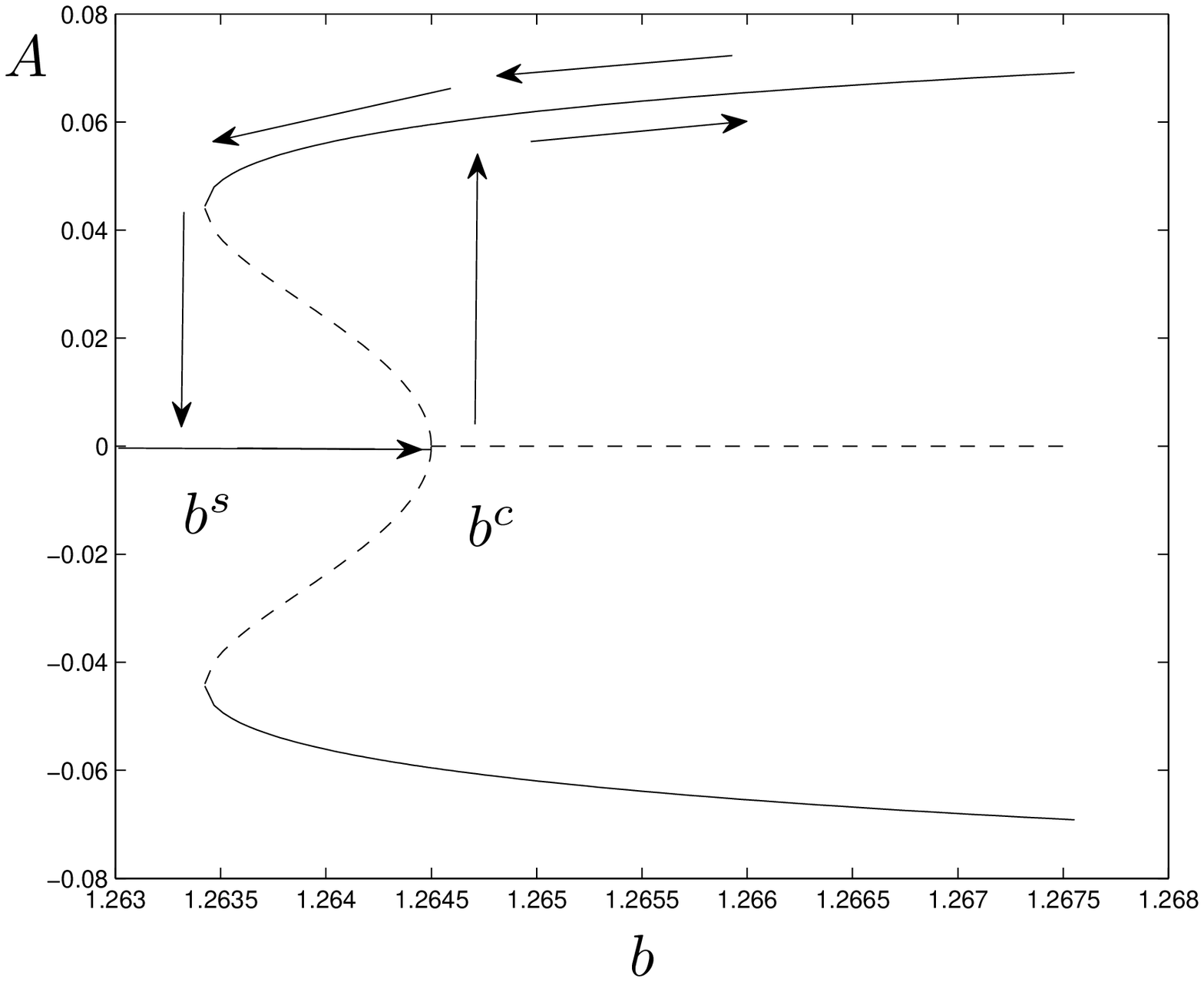}}
\subfigure[]{\epsfxsize=6.5cm\epsfysize=5cm\epsfbox{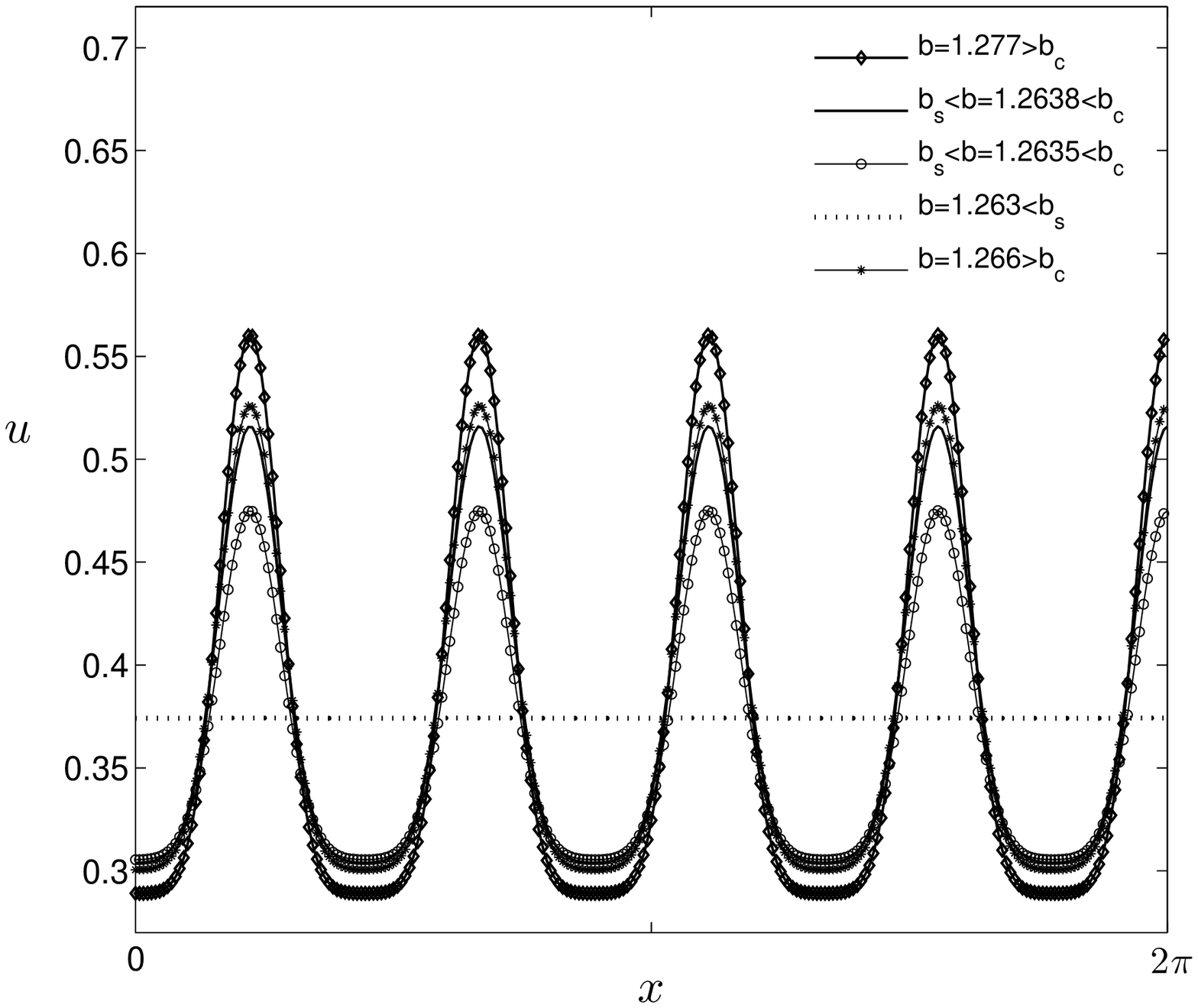}}
\end{center}
\caption{\label{fig3_3} (a) The bifurcation diagram in the subcritical case. (b) A hysteresis cycle and the corresponding pattern
evolution in the subcritical case. The bifurcation value $b$ varies following the order given into the legend. The parameters are $\Gamma= 150$,
$Q^2=0.14$, $\eta^2=0.36$, chosen in the subcritical region, at the point marked with an asterisk both in Fig.\ref{fig2_1} and in Fig.\ref{fig3_1}-(a). In this case $b^c\approx 1.2645$, $b^s=1.2634$.}
\end{figure*}
%%%%%%%%%%%%%%%%%%%%%%%%%%%%%%%%%%%%%%%%%%%%%%%%%%%%%%%%%%%%%%%%%%%%%%%%%%%%%%%%%%%%%%%%%%%%%%%
In the subcritical region shadowed in dark grey in Fig.\ref{fig2_1}, the Stuart-Landau equation \eqref{3.11} does not admit any stable equilibrium and it is not able to capture the amplitude of the pattern. We therefore need to push the weakly nonlinear analysis at a higher order
(see \cite{BMS09} and references therein).
The compatibility condition imposed at $O(\varepsilon^5)$
leads to the following quintic Stuart-Landau equation
for the amplitude $A$:
\begin{equation}\label{quintic_SL}
\frac{d A}{d
T}=\bar{\sigma}A-\bar{L}A^3+\bar{R}A^5\, ,
\end{equation}
where the coefficients $\bar{\sigma}$,  $\bar{L}$ and $\bar{R}$
appearing in \eqref{quintic_SL} are obtained in terms of the parameters of the original system \eqref{or_syst}.
The explicit expressions are too involved and are not reported here; however we notice that
$\bar{\sigma}$ and $\bar{L}$ are $O(\varepsilon^2)$ perturbation of the coefficients of the cubic
Stuart-Landau equation, while $\bar{R}=O(\varepsilon^2)$; this leads to $A=O(\varepsilon^{-1})$
consistently with the reported bifurcation diagram.
When $\bar{\sigma}>0$, $\bar{L}<0$ and
$\bar{R}<0$, the equation \eqref{quintic_SL} admits two symmetric real stable equilibria, corresponding to
the asymptotic values of the amplitude $A$.

On the left of Fig.\ref{fig3_3}, the  bifurcation diagram
is shown for the values of the parameters chosen in the subcritical region.
When $b^s<b<b^c$ both the origin and two large
amplitude branches are stable, indicating the possibility of
hysteresis as $b$ is varied.
On the right of Fig.\ref{fig3_3}, it is shown how the pattern forms
starting with a value of $b>b^c$, as the solution of the equation \eqref{quintic_SL} jumps to the large amplitude stable branch. Moreover, decreasing $b$, with $b^s<b<b^c$ this pattern solution persists; it disappears, reaching the constant steady state, only with a further decrease
of $b$ below $b^s$, as the solution of \eqref{quintic_SL} jumps to the origin. The pattern forms again by increasing the
parameter $b$ above $b^c$.
Finally, we notice that in the subcritical case the amplitude of the pattern is relatively
insensitive to the size of the bifurcation parameter.
\section{Two dimensional domain\label{Sec4}}
\setcounter{equation}{0}
In this section we shall investigate the pattern appearance for the reaction-diffusion
system \eqref{or_syst} in a two-dimensional domain (here $\zeta\in \Omega\subseteq\mathbb{R}^2$).
Notice that the critical value for the bifurcation parameter and the critical wavenumber do not depend on the geometry of the domain and
they are still computed via linear stability analysis as in Section \ref{Sec2}.
\subsection{Rectangular domain}
Let ${\zeta}\equiv(x,y)\in \Omega$, with $\Omega=[0,L_x]\times[0,L_y]$.
The solutions to the linearized system associated to \eqref{or_syst}
with Neumann boundary conditions are:
\begin{eqnarray}\label{sol2d}
&&{\bf w}=\sum_{p,q\in \mathbb{N}}\mathbf{f}_{pq}\,e^{\sigma ({k}_{pq}^2)\,t}\,
\cos\left(\phi \, x\right)\cos\left(\psi\,y\right),\\ \label{k2d}
&&k_{pq}^2=\phi^2+\psi^2, \ {\rm where}\ \ \phi\equiv \frac{p\pi}{L_x},\ \
\psi\equiv  \frac{q\pi}{L_y},
\end{eqnarray}
where $\mathbf{f}_{pq}$ are the Fourier coefficients of the initial conditions and $\sigma(k_{pq}^2)$ are computed via the dispersion relation \eqref{2.4}.
The range of the unstable wavenumbers of allowable patterns strictly depends on the domain geometry and the boundary conditions.
Being the domain finite, to see a pattern emerging as $t$ increases, there should exist at least a
mode pair $(p,q)$ such that:
\begin{equation}\label{con2d}
\begin{split}
&k_1^2<k^2\equiv \phi^2+\psi^2<k_2^2\ \ {\rm and}\ \ \sigma(k^2)>0,\\
&{\rm where}\ \ \phi\equiv \frac{p\pi}{L_x}\ \ {\rm and}\ \
\psi\equiv  \frac{q\pi}{L_y},
\end{split}
\end{equation}
i.e. for $b>b^c$ and $\Gamma$ sufficiently large (as the unstable wavenumbers
$k_1^2$ and $k_2^2$ are proportional to $\Gamma$).
Our analysis will be restricted to the case when only
one admissible unstable eigenvalue, here denoted with $\bar{k}_c$,
falls within the band $(k_1, k_2)$.
Given $\bar{k}_c\in [k_1,k_2]$, the degeneracy phenomenon can occur: one, two or more pairs $(p,q)$  may exist such that
$\bar{k}_c^2=\phi^2+\psi^2$ and the corresponding eigenvalue $\sigma$ will have single, double or higher multiplicity giving rise
to different types of linear patterns.

The weakly nonlinear analysis can be once again carried out to obtain
the equations which rule the evolution of the pattern amplitude near the threshold.
The solution of the linear problem as in \eqref{sequence_1}, satisfying the Neumann
boundary conditions, is given by:
\begin{equation}\label{solWNL2d}
{\bf w}_1=\sum_{i=1}^r A_i(T_1, T_2)\bfrho\cos(\phi_i x)\cos(\psi_i
y)\, ,
\end{equation}
where $r$ is the multiplicity of the eigenvalue, %$k$ is given in \eqref{con2d},
$A_i$ are the slowly varying amplitudes (still arbitrary at this level) and $\bfrho \in {\rm Ker}(\Gamma K^{b^c}-\bar{k}_c^2D^{b^c})$. We shall show the
types of supported patterns when the multiplicity is $r=1$ or $r=2$.
\subsubsection{Simple eigenvalue $r=1$}\label{simple}
\begin{figure}[h]
\begin{center}
\subfigure[] {\epsfxsize=5.66cm \epsfbox{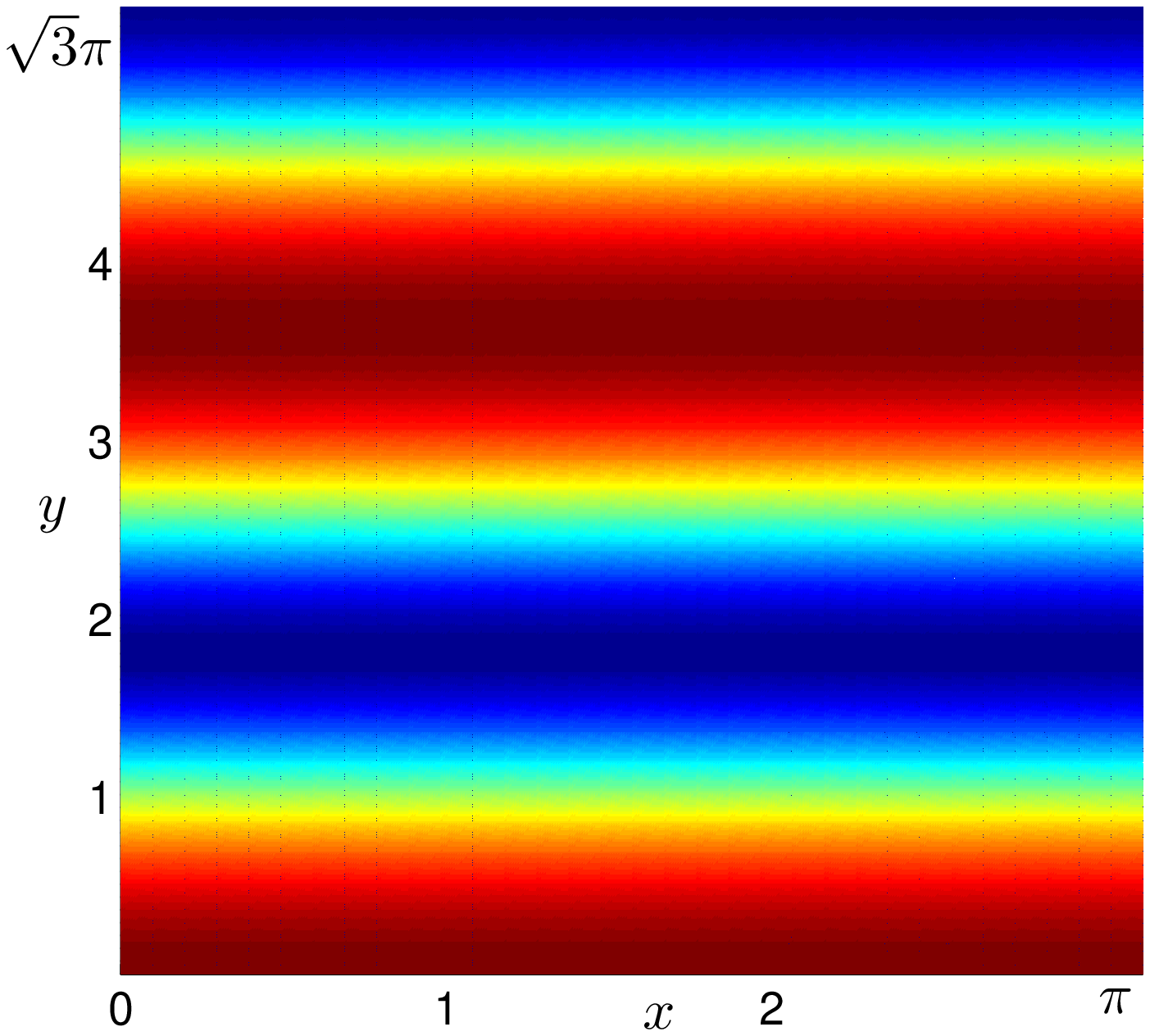}}
\subfigure[] {\epsfxsize=5.65cm \epsfbox{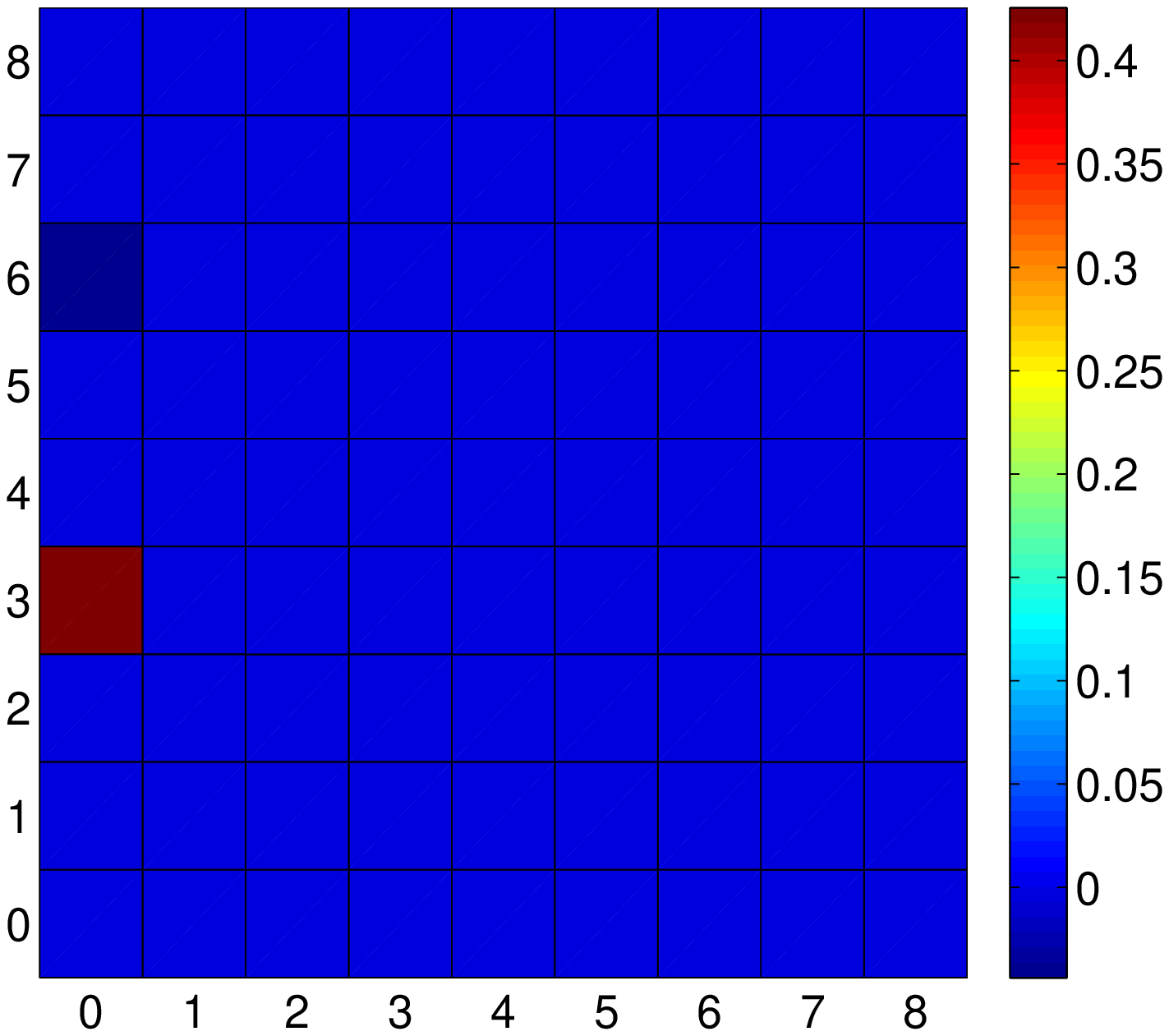}}
\end{center}
\caption{\label{fig4_1}(a) The species $u$. (b) Spectrum of the numerical solution. The parameters are  $\Gamma= 8$, $m=n=1$, $Q^2=3$, $\eta^2=0.36$, $b=b^c(1+\varepsilon^2)$, where $b^c=5.3028$ and $\varepsilon= 0.05$.}
\end{figure}

When $r=1$ the multiple scales method strictly follows the analysis given in Section \ref{Sec3}.
The amplitude equation, still recovered at $O(\varepsilon^3)$, is the Stuart-Landau equation \eqref{3.11} and
the emerging solution of the reaction-diffusion
system \eqref{or_syst} in the supercritical case is given by:
\begin{equation}\label{sol_m1}
\mathbf{w}=\varepsilon \bfrho A_{\infty} \cos(\phi x)\cos(\psi y)+O(\varepsilon^2),
\end{equation}
where $A_{\infty}$ is the stable stationary state of the Stuart-Landau equation \eqref{3.11}.
These solutions are rhombic spatial patterns (see \cite{CMM97}), whose special cases are the rolls (when $\phi$ or $\psi$ is zero) or the squares (when $\phi=\psi$).  The numerical solution, obtained via spectral methods, of the system \eqref{or_syst}, starting from
an initial datum which is a random periodic perturbation about the steady state
$(\bar{u},\bar{v})$, stabilizes to the roll pattern shown
in Fig.\ref{fig4_1}. The system parameters are chosen in such a way that, in the rectangular domain $L_x=\pi$ and $L_y=\sqrt{3}\pi$, only the most unstable
mode $\bar{k}_c^2=3$ satisfies the condition \eqref{con2d} and the corresponding eigenvalue
is single, as the uniform steady state is linearly unstable to the unique mode pair $(p,q)=(0,3)$.
For a better presentation of the results the amplitude of the zero mode (corresponding to the equilibrium solution) has been set equal to zero into the figures representing the spectrum of the solution. We have verified that the
error in predicting the amplitude of the pattern using \eqref{sol_m1} is $O(\varepsilon^2)$ (see the presence of
the subharmonic $(0, 6)$ in  Fig.\ref{fig4_1}(b) which can be estimated including into the approximated solution \eqref{sol_m1} also the terms at $O(\varepsilon^2)$).

\begin{figure}[h]
\begin{center}
\subfigure[] {\epsfxsize=5.6cm\epsfbox{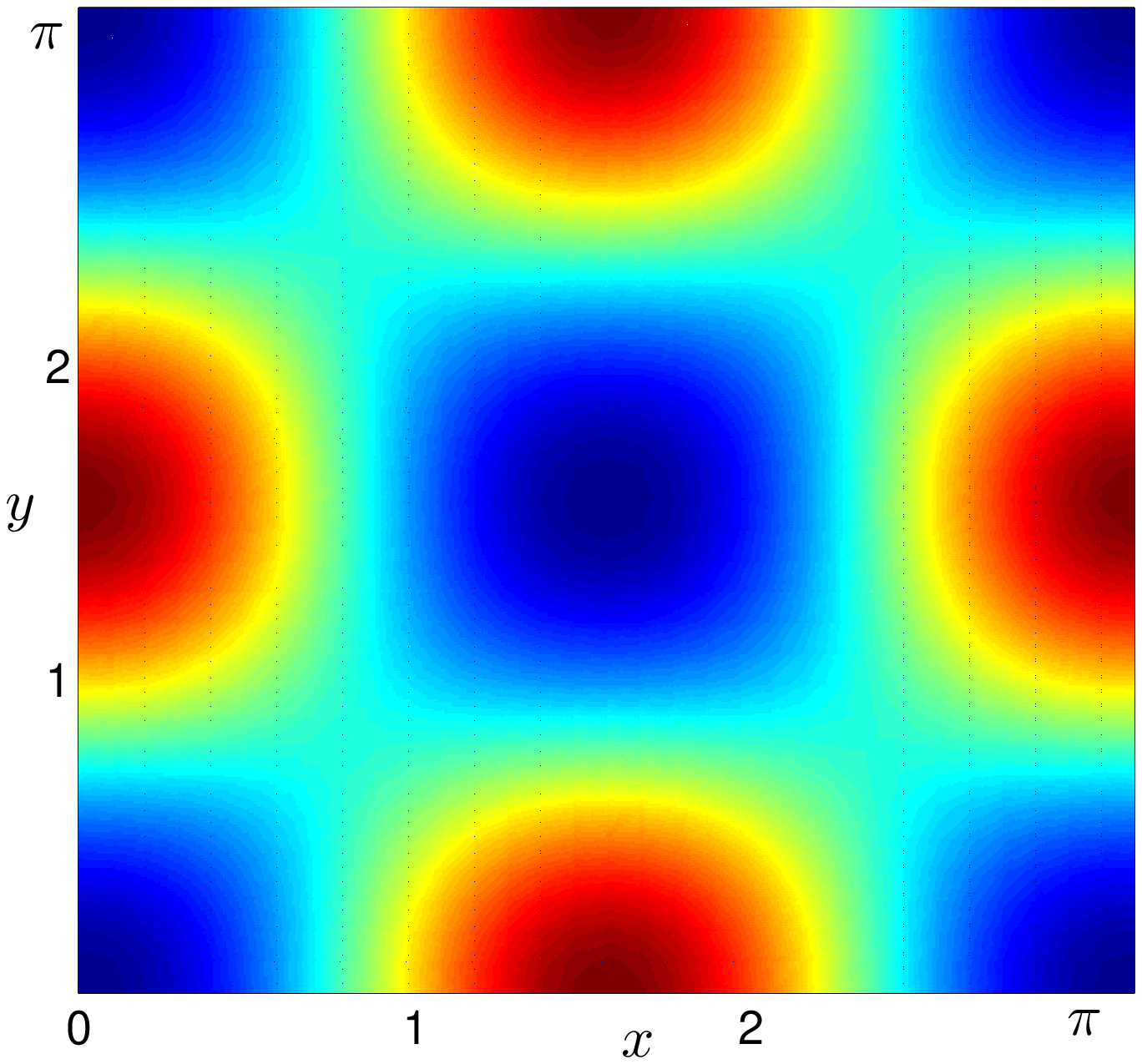}}
\subfigure[] {\epsfxsize=5.6cm \epsfbox{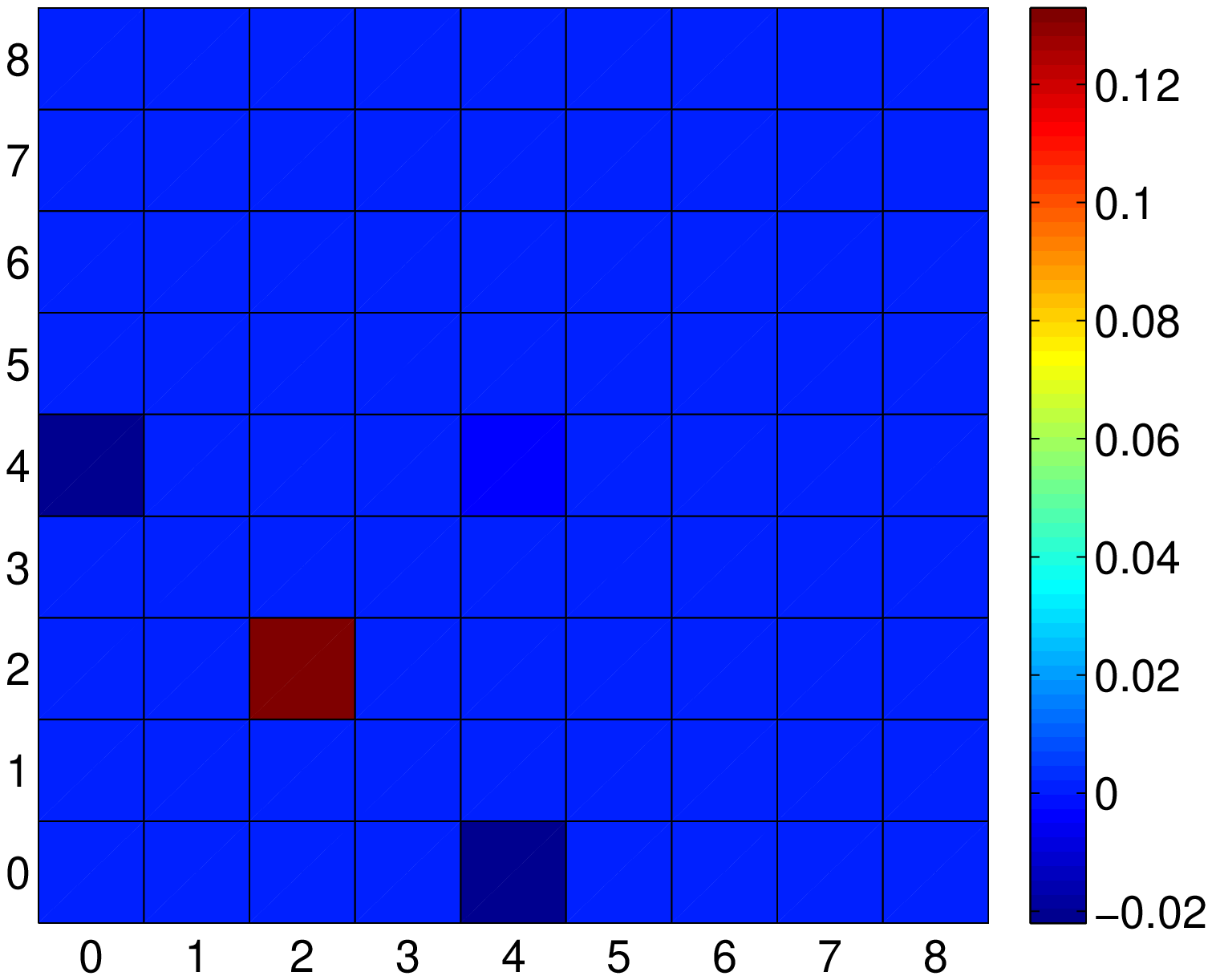}}
\end{center}
\caption{\label{fig4_2}(a) The species $u$. (b) Spectrum of the numerical solution. The parameters are  $\Gamma= 30.3$, $m=1$, $n=2$, $Q^2=3.5$, $\eta^2=0.81$, $b=b^c(1+\varepsilon^2)$, where $b^c=3.9542$ and $\varepsilon= 0.02$.
}
\end{figure}

In the square domain with $L_x=L_y=\pi$, we have picked the parameters values
as in the caption of Fig.\ref{fig4_2}, such that the conditions in \eqref{con2d} are satisfied
by the unique discrete unstable mode $\bar{k}_c^2=8$ and the mode pair $(p,q)=(2,2)$.
From an initial condition which is a random periodic perturbation about $(\bar{u},\bar{v})$, the numerical solution of the system \eqref{or_syst} stabilizes to
the square pattern given in Fig.\ref{fig4_2}, in agreement with the expected solution \eqref{sol_m1}. The subharmonics $(4,0)$, $(0,4)$ and $(4,4)$ in Fig.\ref{fig4_2}(b) can be predicted
via the weakly nonlinear approximation \eqref{sol_m1} up to  $O(\varepsilon^2)$.

\subsubsection{Double eigenvalue $r=2$, no-resonance condition holds}\label{double_nores}
Let us assume that the multiplicity of the eigenvalue is $r=2$ and the following no-resonance condition
holds:
\begin{eqnarray}
 \phi_i+\phi_j\neq \phi_j   \quad  &\mbox{or}&
\quad\psi_i-\psi_j\neq \psi_j \nonumber \\
 & \mbox{and}&  \label{4.13} \\
\phi_i-\phi_j\neq\phi_j \quad  &\mbox{or}& \quad
\psi_i+\psi_j\neq\psi_j   \nonumber
\end{eqnarray}
with $i,j=1,2$ and $i\neq j$. Also in this case the weakly nonlinear analysis is performed up to
$O(\varepsilon^3)$ and the solvability condition for the equation \eqref{sequence_3} leads to
the following two coupled Landau equations for the amplitudes $A_1$ and $A_2$:
\begin{subequations}\label{4.19}
\begin{eqnarray}\label{4.19a}
\frac{dA_1}{dT_2}&=&\sigma A_1-L_1 A_1^3+R_1 A_1 A_2^2,\\
\frac{dA_2}{dT_2}&=&\sigma A_2-L_2 A_2^3+R_2 A_1^2\
A_2.\label{4.19b}
\end{eqnarray}
\end{subequations}
In the supercritical case, when the system \eqref{4.19} admits at least one stable equilibrium $(A_{1\infty}, A_{2\infty})$, the emerging asymptotic
solution of the reaction-diffusion system \eqref{or_syst} at the leading order is
approximated by:
\begin{equation}\label{m2nores}
\mathbf{w}=\varepsilon \bfrho \sum_{i=1}^2 A_{i\infty}\cos(\phi_i x)\cos(\psi_i y)+O(\varepsilon^2).
\end{equation}
When $A_{1\infty}$ or $A_{2\infty}$ is zero, the solutions in \eqref{m2nores} are the rhombic spatial patterns
described in Section \ref{simple}. When both $A_{i\infty}\neq 0,\ \ i=1,2$, more complex structures arise due to the interaction
of different modes $\phi_i, \psi_i$, the so-called mixed-mode patterns.
The complete classification of the equilibrium points of the system \eqref{4.19} via linear stability analysis is given
in \cite{GLS13}.
\begin{figure}[h]
\begin{center}
\subfigure[] {\epsfxsize=5.65cm \epsfbox{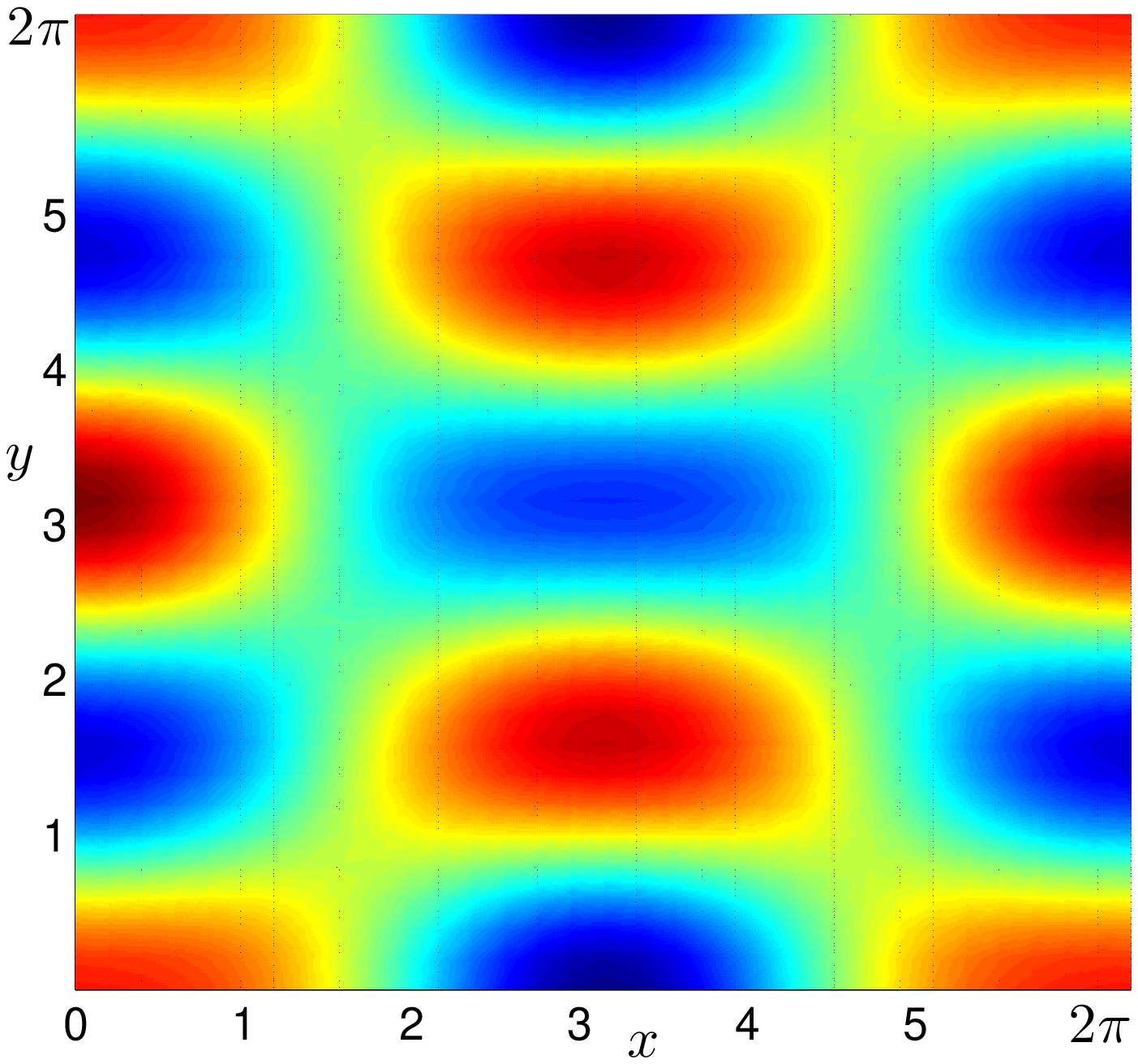} }
\subfigure[] {\epsfxsize=5.65cm \epsfbox{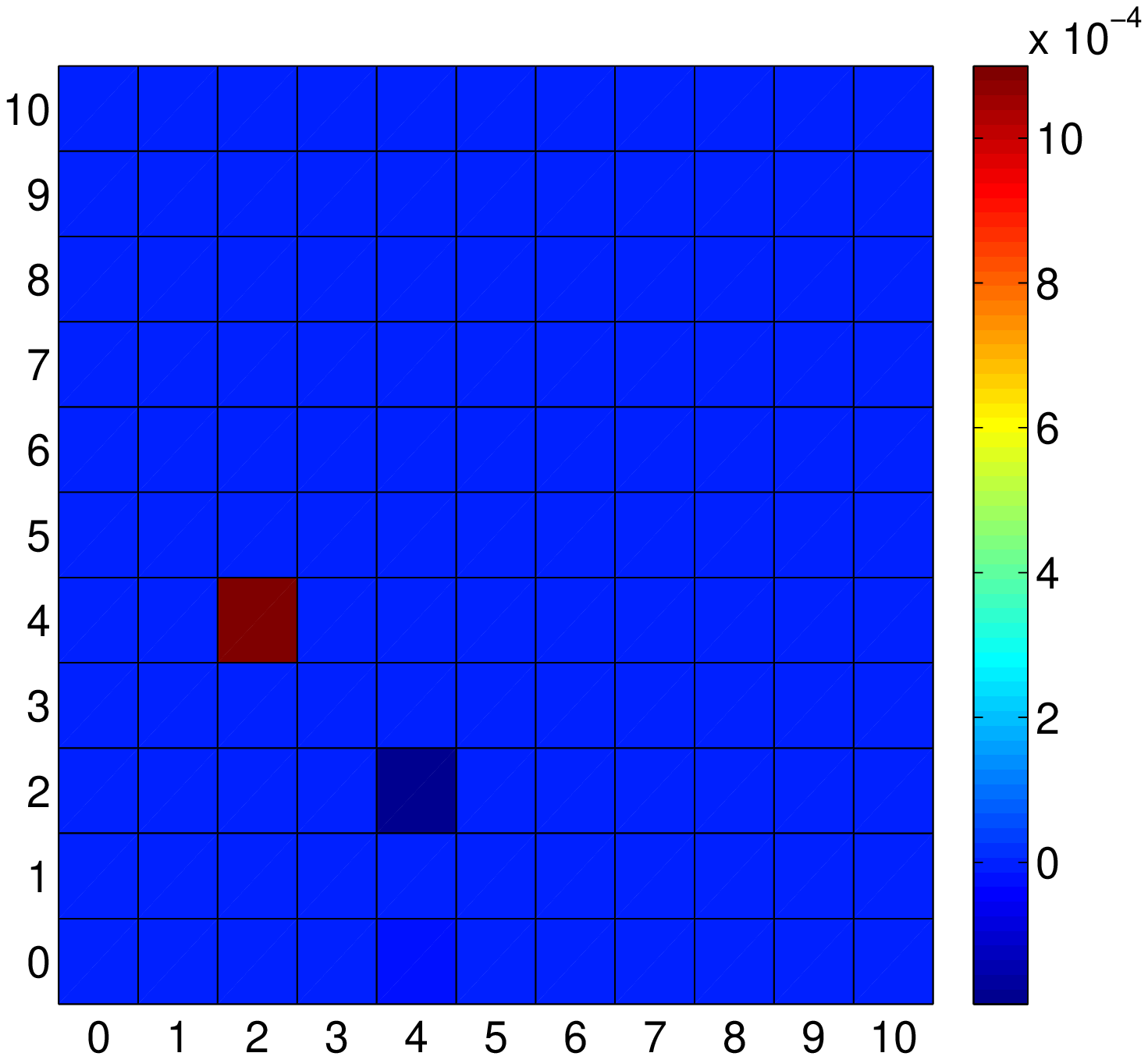}}
\end{center}
\caption{\label{fig4_3} (a) The species $u$. (b) Spectrum of the numerical solution. The parameters are  $\Gamma= 11.93$, $m=n=1$, $Q^2=8$, $\eta^2=0.36$, $b=b^c(1+\varepsilon^2)$, where $b^c=11.3722$ and $\varepsilon= 0.03$.}
\end{figure}

Let us consider a domain with dimensions $L_x=L_y={2}\pi$
and choose the parameter values as in the caption of Fig.\ref{fig4_3} in such a way that only the most unstable discrete
mode $\bar{k}_c^2=5$ falls within the band of unstable modes allowed
by the boundary conditions. In this case the eigenvalue is double, as
the uniform steady state is linearly unstable to the two mode pairs
$(2,4)$ and $(4,2)$. With this choice of the parameters the system \eqref{4.19} admits only one stable equilibrium
$(A_{1\infty}, A_{2\infty})$ with both nonzero coordinates and the expected solution in \eqref{m2nores} agrees with
the numerical asymptotic solution of the system \eqref{or_syst}, having as initial datum a random periodic perturbation about the steady state $(\bar{u},\bar{v})$, shown in
Fig.\eqref{fig4_3}.

\subsubsection{Double eigenvalue $r=2$, resonance condition holds}\label{double_res}

Let the multiplicity of the eigenvalue be $r=2$ and the resonance
condition be satisfied as follows:
\begin{eqnarray}
 \phi_i+\phi_j = \phi_j   \quad  &\mbox{and}&
\quad\psi_i-\psi_j = \psi_j \nonumber \\
 & \mbox{or}&  \label{res_si} \\
\phi_i-\phi_j  =\phi_j \quad  &\mbox{and}& \quad
\psi_i+\psi_j = \psi_j   \nonumber
\end{eqnarray}
with $i, j=1, 2$  and $i\neq j$.
Assuming, without loss of generality, that the second condition in \eqref{res_si} holds with $i=2$ and $j=1$, and
taking into account the relation in \eqref{con2d}, it  follows  that $\phi_2=2\phi_1$, $\psi_2=0$,
$\psi_1= \sqrt{3}\phi_1$, $\phi_1=  \kcb/2$ and $L_y=\sqrt{3}L_x$.
In this case the secular terms appear at $O(\varepsilon^2)$ in \eqref{sequence_2};
however the amplitude equations one derives  imposing the solvability condition
do not admit stable equilibrium in any parameter regimes.
Therefore the asymptotic analysis has to be pushed to
higher order, see \cite{BMS09,GLS13}.
By the Fredholm alternative for the equation \eqref{sequence_3}, at $O(\varepsilon^3)$ one finds the following system
for the amplitudes $A_1$ and $A_2$:
\begin{equation}\label{4.32}
\begin{split}
\frac{dA_1}{dT}=&\,\sigma_1 A_1-L_1 A_1A_2+R_1 A_1 A_2^2+S_1A_1^3,\\
\frac{dA_2}{dT}=&\,\sigma_2 A_2-L_2A_1^2+R_2 A_1^2\, A_2
+S_2A_2^3,
\end{split}
\end{equation}
where ${\sigma_i}$ and ${L_i}$ are $O(\varepsilon^2)$ perturbation of the coefficients of the amplitude equations found
at $O(\varepsilon^2)$, while ${R_i}$ and $S_i$ are $O(\varepsilon^2)$.
At the leading order, the emerging asymptotic solution of the system \eqref{or_syst} is approximated by:
\begin{equation}
\begin{split}\label{4.15}
\mathbf{w}=\varepsilon \bfrho (&\,A_{1\infty}\cos(\phi_1 x)\cos(\psi_1 y)\\
&\,+A_{2\infty}\cos(\phi_2 x)\cos(\psi_2 y))+O(\varepsilon^2),
\end{split}
\end{equation}
where $(A_{1\infty}, A_{2\infty})$ is a stable state of the system \eqref{4.32}. The possible stationary states of the system \eqref{4.32} are $R^{\pm}\equiv(0, \pm\sqrt{-{{\sigma}_2}/{{S}_2}})$ and the six roots $H^{\pm}_i\equiv(A_{1i}^\pm, A_{2i}),\ i=1,2,3,$ of the following system:
\begin{equation}\nonumber
\left\{
\begin{split}
A_2^3(S_1S_2&-R_1R_2)+A_2^2({L}_1R_2+
{L}_2R_1)+\\
&A_2\left(S_1{\sigma}_2-{L}_1{L}_2-R_2{\sigma}_1\right)+{L}_2{\sigma}_1=0,\\
A_1^2=\displaystyle\frac{1}{S_1}&\left(-R_1A_2^2\right.+\left.{L}_1A_2-{\sigma}_1 \right).
\end{split}
\right.
\end{equation}
When $R^{\pm}$ or $H^{\pm}_i$ exist real and stable, the corresponding solution \eqref{4.15} is respectively a roll or a hexagonal pattern.
%
%%%%%%%%%%%%%%%%%%%%%%%%%%%%%%%%%%%Inserimento%%%%%%%%%%%%%%%%%%%%%%%%%%%%%%%%%%%%%%%%%%%%
\begin{figure*}[!]
\begin{center}
\subfigure[] {\epsfxsize=2.0 in \epsfbox{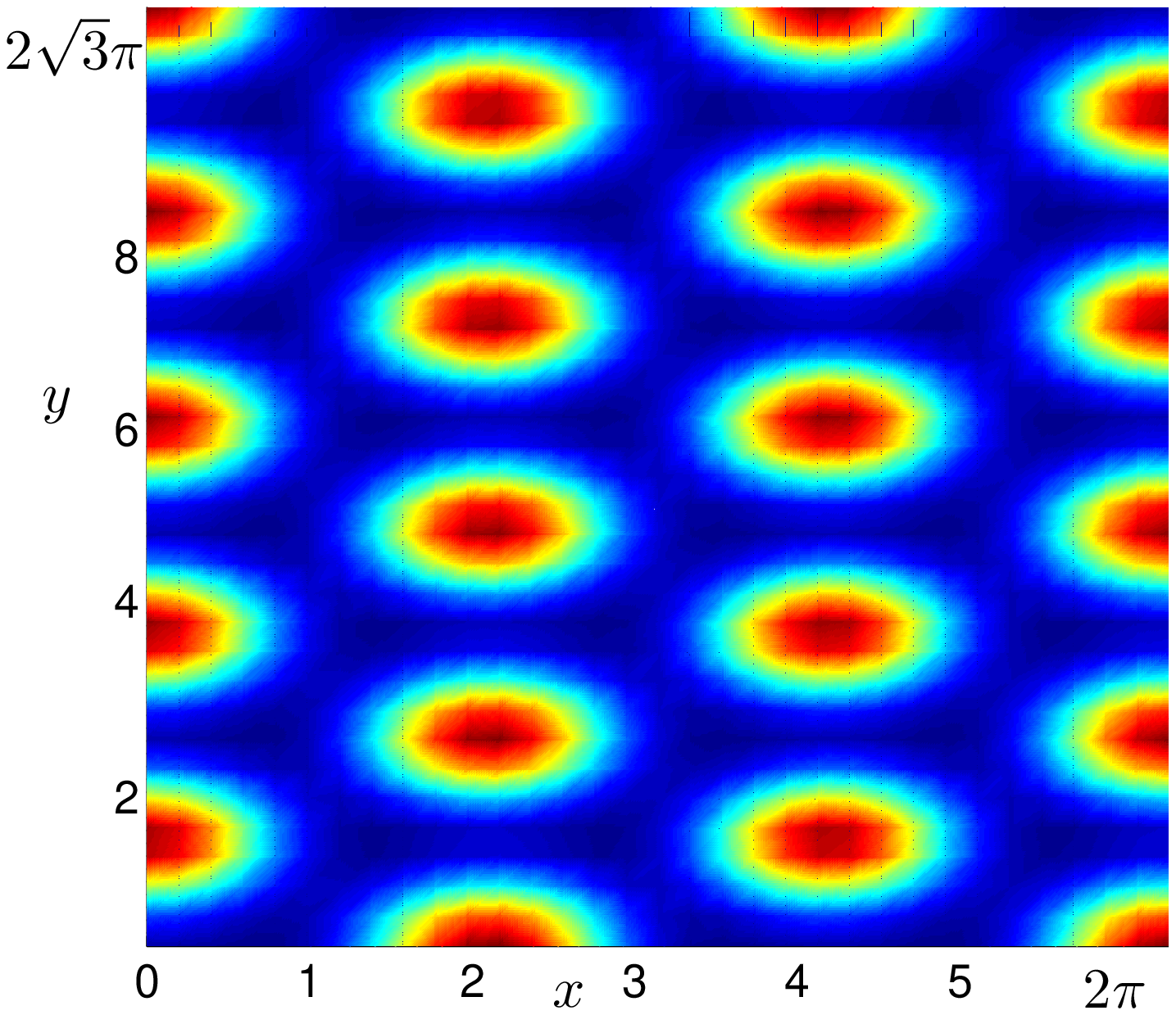}}
\subfigure[] {\epsfxsize=2.0 in \epsfbox{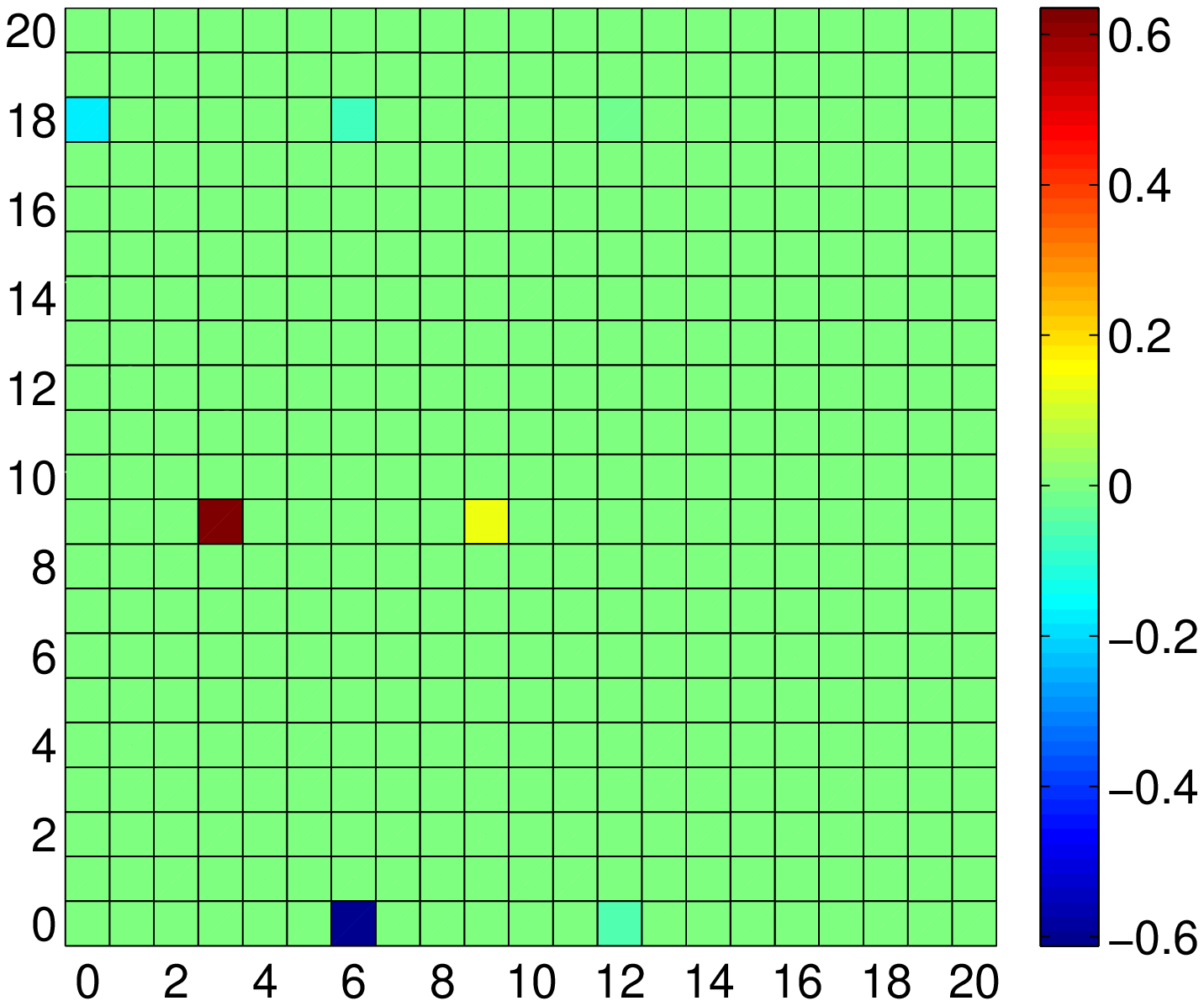}}
\subfigure[] {\epsfxsize=2.0 in \epsfxsize=1.9 in\epsfbox{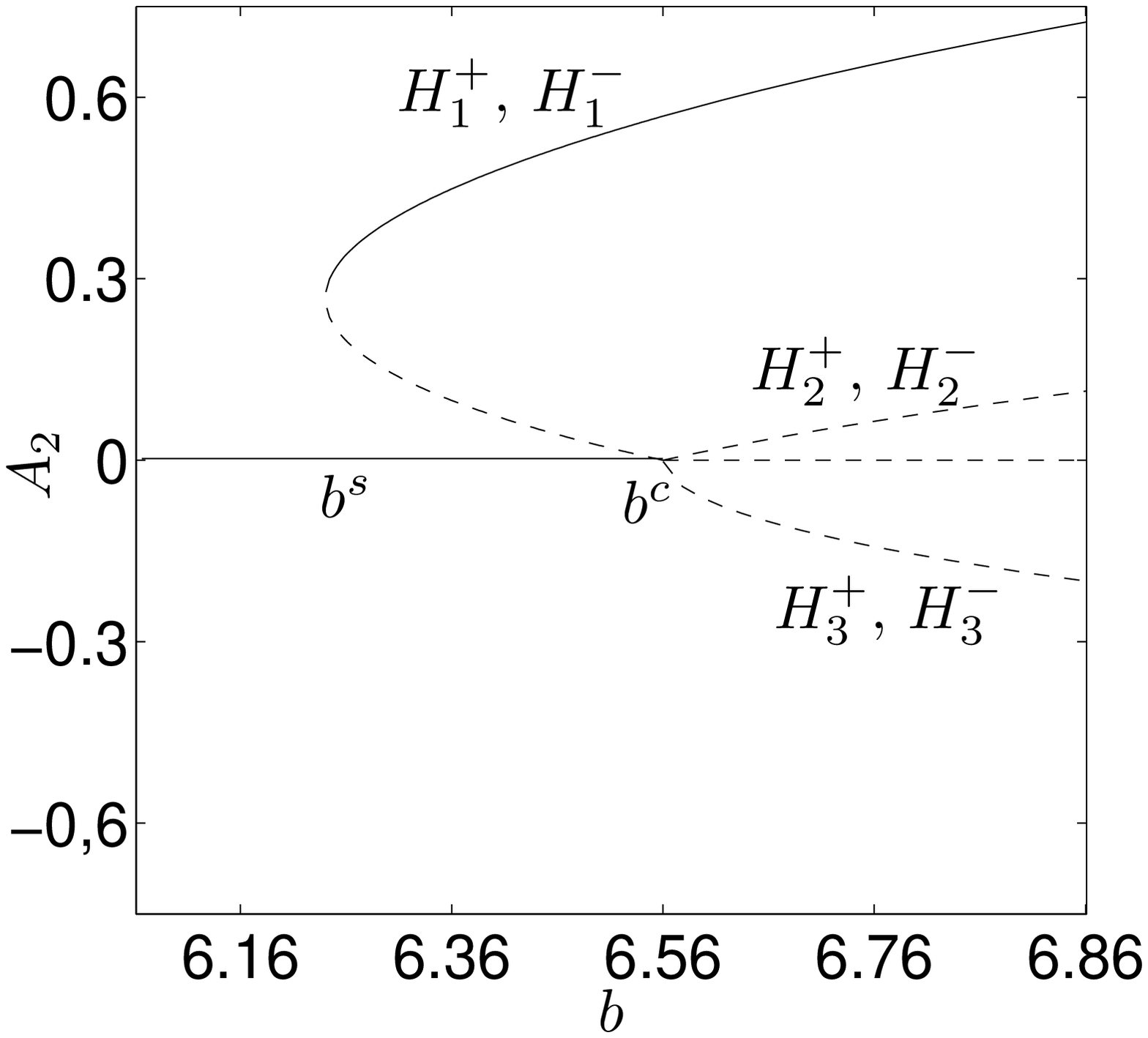}}
\end{center}
\caption{\label{fig4_4}(a) The species $u$. (b) Spectrum of the numerical solution. (c) The bifurcation diagram. The parameters are  $\Gamma= 23.054$, $m=n=1$, $Q^2=4$, $\eta^2=0.3025$, $b=b^c(1+\varepsilon^2)$, where $b^c=6.5615$ and $\varepsilon= 0.01$.}
\end{figure*}
%%%%%%%%%%%%%%%%%%%%%%%%%%%%%%%%%%%%%%%%%%%%%%%%%%%%%%%%%%%%%%%%%%%%%%%%%%%%%%%%%%%%%%%%%%%%%%%%
In Fig.\ref{fig4_4} we show the hexagonal pattern which forms starting from
an initial datum which is a random periodic perturbation about the steady state
$(\bar{u},\bar{v})$.
For the parameters chosen as in the caption of Fig.\ref{fig4_4}, only the mode $\bar{k}_c^2=9$ is admitted by the boundary conditions
in the rectangular domain with $L_x=2\pi$ and $L_y=2\sqrt{3}\pi$. The eigenvalue predicted by the linear analysis
is double as both the two pairs $(3,9)$ and $(6,0)$ satisfy the conditions in \eqref{con2d}.
The weakly nonlinear analysis predicts that only the states $H^{\pm}_1$ are stable
(as shown in the bifurcation diagram in Fig.\ref{fig4_4}(c)).
The form of the pattern emerging from  a numerical simulation of the full system, see Fig.\ref{fig4_4}(a), is qualitatively well captured
by the  hexagonal pattern  \eqref{4.15} predicted by the WNL analysis,
which however underestimates the sub-harmonics shown in Fig.\ref{fig4_4}(b),  as it is usual for subcritical cases.

\subsection{Target pattern with radial symmetry}
\begin{figure}[h]
\begin{center}
\subfigure[] {\epsfxsize=5.65cm \epsfbox{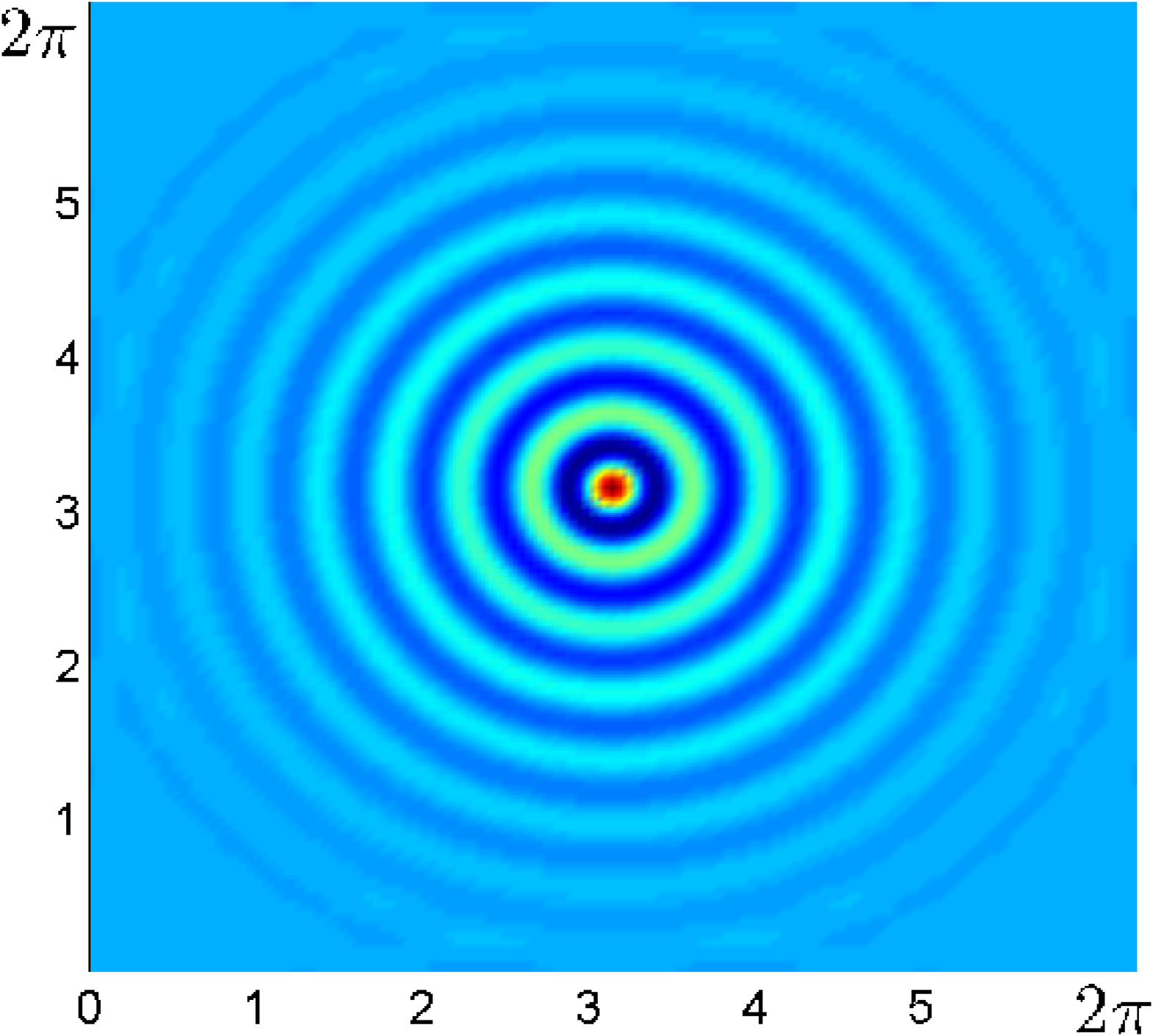}}
\subfigure[] {\epsfxsize=5.65cm \epsfbox{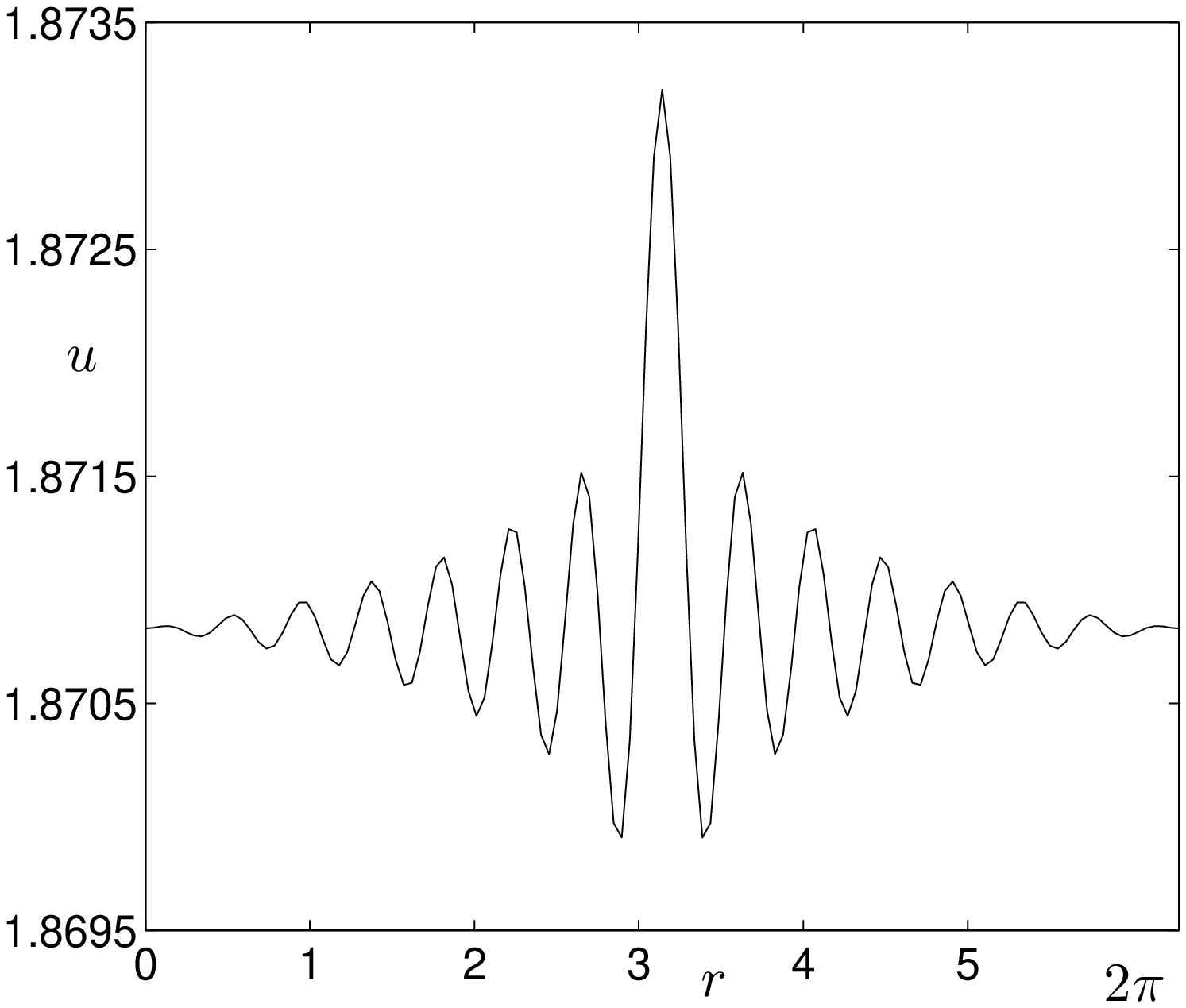}}
\end{center}
\caption{\label{fig11} (Color online) (a) Target pattern. (b) The cross-section.}
\end{figure}

Giving a small radially symmetric perturbation of the uniform equilibrium at the
center of a square domain,
the emerging solution of the system \eqref{or_syst} is the axisymmetric pattern
shown in Fig.\ref{fig11}.
%It is a target pattern consisting of concentric waves that are periodically emitted
%from the small central core.
%
It is a typical target pattern showing a larger amplitude at the center, see
Fig.\ref{fig11}(b).
The weakly nonlinear analysis can be performed to recover
the Ginzburg-Landau equation which captures
the amplitude of the fluctuations of the target pattern close to the
threshold \cite{SM06,WSBR04}. Let $r$ be the spatial radial coordinate, the
amplitude of the pattern
depends on the slow spatial scale $R=\varepsilon r$.
Away from the core the curvature effects can
be neglected and the following Ginzburg-Landau equation
is easily derived following the procedure as
in Section \ref{Sec3}:
\begin{equation}\label{GLtarget}
\frac{\partial A}{\partial T}= \nu
\left(\frac{\partial^2 {A}}{\partial R^2}
+\frac{1}{R}\frac{\partial A}{\partial R}
-\frac{A}{4 R^2}\right) + \sigma A -L A^3 \; .
\end{equation}
The envelope evolution of the outer solution $\textbf{w}$ is therefore approximated by:
\begin{equation}\label{pertTarget}
\textbf{w}_O=
{\varepsilon}\,
{A(R,T)}\,
\textbf{w}_{21} {\cos{( k_c
\bar{r})}}
+O({\varepsilon^2}),
\end{equation}
where $\bar{r}=r-\pi/4$ and $\textbf{w}_{21}$ is the solution of a linear system as
in \eqref{lin_sisGL}.
Let us rewrite the amplitude equation \eqref{GLtarget} in terms of the variable
$\mathcal{A}=A R^{1/2}$ as it simplifies to
the following form:
\begin{equation}\label{GLtar2}
\frac{\partial \mathcal{A}}{\partial T}=\partial_{RR}\mathcal{A}+\sigma
\mathcal{A}-L \frac{\mathcal{A}^3}{R} \; .
\end{equation}
The amplitude equation \eqref{GLtar2} does not
hold close to the center of the target pattern where the
curvature terms cannot be
neglected. Taking into account the curvature terms, through a linear analysis as in
\cite{N91} we get
the following inner solution (close to the core of the pattern) depending on the
spatial radial coordinate $r$:
\begin{equation}\label{star}
\textbf{w}_I
= C \textbf{w}_{21} J_0(k_c r),
\end{equation}
where $J_0$ is the zeroth order Bessel function of first kind and $C$ is a constant.

The solution of the equation \eqref{GLtar2} when $R\rightarrow 0$ should match with
the solution \eqref{star}
when $r\rightarrow \infty$.
The behavior of the solution of
the equation \eqref{GLtar2} when $R\rightarrow 0$ is the following:
\begin{equation}
\mathcal{A}\approx a+ b R+a|a|^2 R \log{R} \; ,
\end{equation}
where $a$ and $b$ are constants, see \cite{S03}.
Using the well known asymptotic
formula for the Bessel function $J_0$, one finds that, when $r\rightarrow \infty$,
the inner solution behaves as:
\begin{equation}\label{bessel}
\textbf{w}_I\approx
\frac{C\textbf{w}_{21}}{\sqrt{\pi k_c\bar{r}}}\cos{(k_c\bar{r})}.
\end{equation}
The matching between the two solutions leads to the constant
$C$ being $O(\ep^{1/2})$. Therefore the solution in the core
$\textbf{w}_I= \ep^{1/2}\ \textbf{w}_{21}\  J_0(k_c r)$
is larger than in the outer
region \cite{PZM85}. This explains the larger amplitude at the center of the
axisymmetric solution of the system \eqref{or_syst}
observed in Fig.\ref{fig11}(b).
\section{Conclusions}
In this paper we have analyzed a two-species reaction-diffusion system
which models the Brusselator dynamics with nonlinear density-dependent diffusion.
We have firstly derived the conditions both
for Turing and oscillatory instabilities, showing that the presence of nonlinear
diffusion extends the range of diffusion coefficients over which Turing patterns can occur, in particular
even when the diffusion coefficient of the activator exceeds that one of the inhibitor.

In one dimensional domain the supercritical or the subcritical character of the Turing bifurcation
has been determined by deriving the amplitude equation for patterns
near the instability threshold via weakly nonlinear analysis. In the subcritical case we have also shown that
the system exhibits hysteresis, as the amplitude equation supports bistability.
Moreover, when the domain is large, we have observed the pattern forming sequentially and invading
the whole domain as a traveling wavefront, whose evolution is governed by the Ginzburg-Landau equation.

In a two dimensional rectangular domain we have observed a rich scenario of diverse patterns, such as
rolls, squares and mixed-mode patterns emerging due to the interaction of different modes.
Among mixed-mode patterns we have also shown the hexagonal patterns, arising when a resonance
condition holds. We have employed the weakly nonlinear analysis to obtain the amplitude equations in each case
and numerical simulations of the reaction-diffusion system exhibit the features
predicted by these amplitude equations.
Finally the analysis has been moved to target pattern with radial symmetry. Since this wave pattern
shows a larger amplitude near the center of its circular profile than in its traveling fluctuations,
we have applied a matching procedure to appropriately approximate the amplitude solution.

Some aspects of the problem remain to be examined.
As the homogeneous state can lose its stability via a Hopf
bifurcation, non-stationary patterns should also develop. The weakly nonlinear analysis can be employed to obtain the amplitude
equations both near the Hopf bifurcation point \cite{IKS00} and next to the codimension-2 Turing-Hopf point here determined \cite{TBMV11}.
This will be the subject of a forthcoming paper.

\section*{Acknowledgements}
The work of the first (GG) and fourth (VS) authors  has been partially supported by GNFM/INdAM through a
\textit{Progetto Giovani} grant.

\bibliographystyle{plain}
\bibliography{pattern}
\end{document}